\documentclass{aa}
\usepackage{graphicx}
\graphicspath{{figures/}}
\usepackage{newtxtext, newtxmath}
\usepackage{siunitx}
\DeclareSymbolFont{upgreek}{U}{eur}{m}{n}
\DeclareMathSymbol{\umu}{0}{upgreek}{"16}
\DeclareSIPrefix{\micro}{\text{\ensuremath{\umu}}}{-6}
\usepackage[colorlinks=true, allcolors=blue]{hyperref}
\begin{document}

    \title{Polarimetric investigation of selected cloud compositions in exoplanetary atmospheres}
    
    
    
    \author{M. Lietzow\and S. Wolf}
    
    
    \institute{Institute of Theoretical Physics and Astrophysics,
              Kiel University, Leibnizstr. 15, 24118 Kiel, Germany\\
              \email{mlietzow@astrophysik.uni-kiel.de}
              }
    
    \date{Received / Accepted }


    \abstract
    {}
    {We investigated the impact of selected cloud condensates in exoplanetary atmospheres on the polarization of scattered stellar radiation.}
    {We considered a selection of 25 cloud condensates that are expected to be present in extrasolar planetary atmospheres.
    Using the three-dimensional Monte Carlo radiative transfer code POLARIS and assuming Mie scattering theory, we calculated and studied the net polarization of scattered radiation as a function of planetary phase angle at optical to near-infrared wavelengths (\SI{0.3}{\um} to \SI{1}{\um}).}
    {In addition to the well-known characteristics in the state of polarization, such as the rainbow determined by the real part of the refractive index, the behavior of the underlying imaginary part of the refractive index causes an increase or decrease in the degree of polarization and a change of sign in the polarization at a characteristic wavelength.
    In contrast to Al$_2$O$_3$ and MgFeSiO$_4$, clouds composed of SiO, MnS, Na$_2$S, or ZnS produce a rapidly decreasing degree of polarization with increasing wavelength in the context of an exoplanetary atmosphere.
    Furthermore, the sign of the polarization changes at a wavelength of about \SI{0.5}{\um} to \SI{0.6}{\um}, depending on the specific cloud condensate.
    The resulting net polarization is mainly positive for cloud compositions with large imaginary parts of the refractive index, such as Fe, FeS, and FeO.
    In addition, for Fe and FeS clouds, the maximum degree of polarization at long wavelengths is shifted to larger phase angles than for FeO.}
    {We found that most of these cloud condensates, such as chlorides, sulfides, or silicates, are distinguishable from each other due to their unique wavelength-dependent complex refractive index.
    In particular, an increase or decrease of the net polarization as a function of wavelength and a change of sign in the polarization at specific wavelengths are important features for characterizing cloud compositions in exoplanetary atmospheres.}


    \maketitle
%
\section{Introduction}

Clouds are an essential part of atmospheres of most planets in the Solar System, such as sulfuric acid clouds on Venus \citep{young1973}, water clouds on Earth, or ammonia, ammonium hydrosulfide, and water-ice clouds on Jupiter or Saturn \citep{atreya1999}.
Cloud particles can form and grow when supersaturation occurs.
If the saturation vapor pressure curve of a cloud species crosses the planetary pressure-temperature profile, the particles condensate and form clouds \citep{sanchez-lavega2004}.
Condensation of particles includes simple phase changes (e.g., water, ammonia) or thermochemical reactions, for example, in the case of ammonium hydrosulfide in the atmosphere of Jupiter \citep{lodders2002}.
In addition to Solar System objects, the existence of clouds is predicted \citep[e.g.,][]{burrows1997, ackerman2001, helling2001, lodders2002, morley2012}.
Their signature has been observed in the atmosphere of brown dwarfs and extrasolar planets \citep[e.g.,][]{charbonneau2002, pont2008, sing2009, demory2011, gibson2013, morley2013}.

Cloud compositions do not only leave an imprint on the transmission spectra from the optical to infrared \citep{morley2014, wakeford2015}, they also affect the planetary albedo and therefore the observed reflected flux as a function of the planetary phase angle \citep[e.g.,][]{seager2000}.
Furthermore, radiation scattered in the planetary atmosphere is usually polarized, whereas the net polarization of inactive solar-type stars is assumed to be negligible \citep{kemp1987, cotton2017}.
Consequently, measuring the scattered polarized flux of a planet at various phase angles and wavelengths allows determining atmospheric properties.
\citet{hansen1974a} demonstrated this method by characterizing the size distribution and real part of the refractive index of the cloud particles in the atmosphere of Venus using polarimetric observations by \citet{lyot1929} and \citet{coffeen1969}.
Thus, polarimetry also represents a promising tool for the characterization of exoplanetary atmospheres.

With modern instruments, such as HIPPI \citep{bailey2015}, HIPPI-2 \citep{bailey2020}, POLISH2 \citep{wiktorowicz2015}, POLLUX \citep{muslimov2018}, or SPHERE \citep{beuzit2019}, exoplanet polarimetry has come into reach.
Various studies of scattered polarized radiation of exoplanets have been made.
Selected examples are predictive studies of the reflected flux and polarization of close-in extrasolar giant planets \citep{seager2000}, cloud-free Rayleigh scattering planetary atmospheres \citep{buenzli2009}, Jupiter-like gas giants with homogeneous \citep{stam2004} and inhomogeneous cloud coverage \citep{karalidi2013}, Earth-like exoplanets with homogeneous \citep{stam2008, karalidi2011} and inhomogeneous cloud coverage \citep{karalidi2012a}, the retrieval of cloud coverage of Earth-like planets \citep{rossi2017}, and the polarization of self-luminous exoplanets \citep{stolker2017}.

The first detection of polarized scattered radiation of an exoplanet was reported by \citet{berdyugina2008}.
The authors were able to obtain polarimetric measurements of the HD~189733 system in B band.
The system harbors a planet with an orbital semimajor axis of only about \SI{0.03}{AU} and a period of about \SI{2.2}{days} \citep{bouchy2005}.
Assuming Rayleigh scattering, the authors concluded that the observed modulation was due to scattered radiation from the atmosphere of HD~189733b.
This first detection was also confirmed by \citet{berdyugina2011}.
New observations of HD~189733 by \citet{wiktorowicz2015} and \citet{bott2016} also reported a linear polarization signal and confirmed the previously reported observations, but with a lower amplitude of polarization variations.
\citet{bailey2021}, however, assumed that the highly variable polarization has its origin in the magnetic activity of the host star.

Another system for which linear polarization was reported by \citet{bott2018} is WASP-18.
This system harbors a planet with an orbital distance of about \SI{0.02}{AU} and a period of \SI{0.94}{days} \citep{hellier2009}.
Although the polarization is dominated by the additional influence of the interstellar medium, the authors set an upper limit of \SI{40}{ppm} on the amplitude of reflected polarized radiation from WASP-18b.

Our study is organized as follows:
In Sect.~\ref{sec:methods} we start with a brief overview of the applied methods.
Our model of an exoplanetary atmosphere is described in Sect.~\ref{sec:model-setup}.
Subsequently, the first goal of this study was to review characteristic properties of the polarization state of the radiation that is scattered once (hereafter: single-scattered) by cloud particles.
For this purpose, in Sect.~\ref{subsec:single-scattering}, we calculate optical properties of selected cloud particle compositions that are expected to condensate in various types of planetary atmospheres, ranging from ice giants to hot Jupiters.
Our second goal was to investigate the scattered polarized flux of cloudy planetary atmospheres with various cloud compositions, which we report in Sect.~\ref{subsec:exoplanet-scattering}.
Therein, we compare the considered cloud compositions, discuss the polarization, and evaluate the feasibility of using characteristic features in the net polarization to distinguish between these cloud particles.
To study the polarization at various phase angles and wavelengths ranging from \SI{0.3}{\um} to \SI{1}{\um}, the publicly available\footnote{\href{http://www1.astrophysik.uni-kiel.de/~polaris/}{\texttt{www1.astrophysik.uni-kiel.de/\textasciitilde{}polaris}}} three-dimensional Monte Carlo radiative transfer code POLARIS \citep{reissl2016} was applied.
It is well tested and has been used to investigate a broad range of astrophysical models such as molecular clouds \citep{reissl2017, pellegrini2020, seifried2020}, Bok globules \citep{brauer2016, zielinski2021}, and protoplanetary disks \citep{heese2020, brunngraeber2020, brunngraeber2021}.
For the purpose of our study, it has been optimized to handle the radiative transfer in planetary atmospheres \citep{lietzow2021}.
Finally, a discussion and a brief study of the observability of these polarization features is given in Sect.~\ref{sec:discussion}.
Our conclusions are summarized in Sect.~\ref{sec:conclusions}.

\section{Methods}
\label{sec:methods}

To calculate the scattered (polarized) flux of an exoplanet, we simulated the radiative transfer in its atmosphere with a Monte Carlo method.
We used the three-dimensional Monte Carlo radiative transfer code POLARIS \citep{reissl2016}, which has recently been optimized to calculate the scattered radiation of planetary atmospheres \citep{lietzow2021}.
Therein, the radiation field is represented by photon packages that are emitted by a spatially extended radiation source, propagate through the three-dimensional model space, are scattered by atmospheric particles, and are finally detected by an observer.

In the past, radiative transfer calculations applied to planetary atmospheres with the Monte Carlo approach have also been performed, for instance, by \citet{buenzli2009}, \citet{garcia-munoz2015}, \citet{emde2016}, \citet{stolker2017}, or \citet{bailey2018}, while \citet{stam2006} or \citet{rossi2018b}, for example, performed calculations based on the adding-doubling method \citep{de-haan1987}.

To describe the state and degree of polarization of the photon packages, we used the Stokes formalism \citep[e.g.,][]{bohren1983}.
Here, each photon package carries a wavelength-dependent Stokes vector $\vec{S} = (I, Q, U, V)^\mathrm{T}$.
Its four components are the total flux $I$, the components $Q$ and $U$, describing the linear polarization, and the circularly polarized flux $V$.
At each interaction of the photon package with an atmospheric particle, the Stokes vector is multiplied by a scattering matrix $\mathbf{F}(\theta, \phi)$ to account for the change in polarization,
\begin{equation}
    \label{eq:stokes-transformation}
    \vec{S} \propto \mathbf{F}(\theta, \phi) \vec{S}',
\end{equation}
where $\vec{S}'$ is the Stokes vector of the incoming radiation, and $\theta$ and $\phi$ are the scattering angles.
When the photon package leaves the atmosphere, the resulting Stokes vector can be measured by an observer.
The degree $P_\mathrm{l}$ and angle $\chi$ of linear polarization is
\begin{equation}
    P_\mathrm{l} = \frac{\sqrt{Q^2 + U^2}}{I},\quad
    \tan( 2\chi ) = \frac{U}{Q}.
\end{equation}

In studies of Solar System objects, a common approach is to express the degree of polarization in a reference system.
In particular, the degree of polarization is defined as
\begin{equation}
    P_\mathrm{s} = P_\mathrm{l} \cos( 2\vartheta_\mathrm{r} ),\quad
    \vartheta_\mathrm{r} = \chi - \left( \psi \pm \frac{\pi}{2} \right),
\end{equation}
where $\psi$ is the position angle of the scattering plane, that is, a plane containing star, planet (or comet), and observer.
The term inside the brackets must satisfy the condition $0 \leq ( \psi \pm \pi/2) \leq \pi$ \citep{chernova1993}.
As a consequence, the degree of polarization has a sign, and we can distinguish between positive and negative polarization, which means that the radiation is polarized perpendicular and parallel to the scattering plane, respectively.
Furthermore, if the planet is symmetric with respect to the scattering plane, it is $U = 0$, and with $\psi = 0$, we define a signed degree of linear polarization such as
\begin{equation}
    \label{eq:signed-linear-polarization}
    P_\mathrm{s} = -\frac{Q}{I}.
\end{equation}
In this study, the total flux only contains the net scattered planetary flux and not the stellar contribution. The planet can therefore be spatially resolved from its star.

In the case of a homogeneous spherical particle, the scattering matrix given in Eq.~\ref{eq:stokes-transformation} is reduced to
\begin{equation}
    \label{eq:scattering-matrix}
    \mathbf{F} =
    \left(\begin{smallmatrix}
    F_{11} & F_{12} & 0 & 0 \\
    F_{12} & F_{11} & 0 & 0 \\
    0 & 0 & F_{33} & F_{34} \\
    0 & 0 & -F_{34} & F_{33}
    \end{smallmatrix}\right).
\end{equation}
The four essential elements depend on the complex refractive index $n_\mathrm{c} = n_\mathrm{r} + \mathrm{i} n_\mathrm{i}$ of the particle and the size parameter $x = 2 \pi r / \lambda$, where $r / \lambda$ is the ratio of particle radius and wavelength.

For a collection of particles with a size distribution $n(r)$, we define an averaged matrix element as
\begin{equation}
    \label{eq:average-scat-matrix}
    \langle F_{ij} \rangle =
    \frac{\int_{r_\mathrm{min}}^{r_\mathrm{max}} n(r) F_{ij}(r)\ \mathrm{d} r}{\int_{r_\mathrm{min}}^{r_\mathrm{max}} n(r)\ \mathrm{d} r}.
\end{equation}
Considering incident unpolarized radiation, the degree of polarization of single-scattered radiation by a collection of particles following Eqs.~\ref{eq:stokes-transformation} and \ref{eq:signed-linear-polarization} becomes
\begin{equation}
    \label{eq:single-scattering-polarization}
    P_\mathrm{s} = -\frac{\langle F_{12} \rangle}{\langle F_{11} \rangle}.
\end{equation}

\section{Model setup}
\label{sec:model-setup}

We used a spherical model space that contains the planet and the illuminating source (star).
The planet is located at the center of coordinate space, and the star is located at a distance $d$ from the planet.
The planetary atmosphere was divided into a logarithmically spaced radial pressure grid, ranging from a top pressure of \SI{e-5}{bar} to a pressure of \SI{100}{bar}.
The atmosphere consisted of molecular hydrogen. The total scattering optical depth of the gaseous layers therefore was approximately 18 at \SI{0.5}{\um}.

Assuming an absorbing surface underneath, the lower boundary represents a thick cloud layer that absorbs all incoming radiation.
However, the impact of the lower boundary of the atmosphere is negligible, especially at shorter wavelengths, since the information of the polarization state is lost at this optical depth \citep{buenzli2009}.
For the gaseous particles, we only considered scattering and ignored absorption (see Sect.~\ref{sec:discussion} for further discussion).
The scattering cross-section, the refractive index, and the scattering matrix were calculated using the formula given by \citet{sneep2005}, \citet{cox2000}, and \citet{hansen1974b}, respectively.
For molecular hydrogen, a depolarization factor of 0.02 was used \citep{hansen1974b}.

In addition to the gaseous layers, we also considered clouds in the atmosphere.
The clouds consist of small spherical particles whose distribution is characterized by a power law with exponential decay, such as
\begin{equation}
    \label{eq:size-distribution}
    n(r) \propto r^{p_1} \mathrm{e}^{-r / p_2},
\end{equation}
where $r$ is the particle radius.
Because of its simplicity, it is commonly applied to describe various types of clouds \citep{hansen1971}.
This size distribution is sometimes also referred to as the standard (gamma) distribution \citep{hansen1974b}.
The two parameters $p_1$ and $p_2$ depend on the effective radius $r_\mathrm{eff}$ and effective variance $v_\mathrm{eff}$ of the distribution
\begin{equation}
    \label{eq:size-distiribution-parameter}
    p_1 = \frac{1 - 3 v_\mathrm{eff}}{v_\mathrm{eff}}, \quad
    p_2 = r_\mathrm{eff} v_\mathrm{eff}.
\end{equation}

In contrast to the gaseous particles, we considered both scattering and absorption by the cloud particles.
The absorption and scattering cross-section as well as the scattering matrix of the cloud particles were determined using the program \emph{miex} \citep{wolf2004}, which is based on the Mie scattering theory \citep{mie1908}.
An underlying assumption of this approach is that the particles are chemically homogeneous spheres.

Although cloud condensates in planetary atmospheres, such as water droplets, water-ice particles or solid particles in general, can be expected not to be perfectly spherical, the assumption of spherical particles can be used to mimic optical properties, such as cross-sections, of nonspherical particles \citep{kitzmann2010}.
However, polarization is much more sensitive to the particle shape.
Linear polarization of randomly oriented nonspherical particles strongly depends on the aspect ratio \citep{mishchenko1994}.
In addition, for size parameters $\gtrsim$5 and large scattering angles, calculations assuming spheres usually do not agree with the measurements of nonspherical particles, but rounded particles are closely approximated by the Mie scattering theory \citep{perry1978}.
Furthermore, while liquid cloud condensates can be approximated by spheres, such as water clouds on Earth \citep{goloub1994} or sulfuric acid clouds on Venus \citep{hansen1974a, garcia-munoz2014}, solid particles, such as ice crystals in cirrus clouds, show different polarization signals compared to liquid water clouds due to their large variability in shape, size, and density \citep{hess1998, goloub2000}.
These ice particles in cirrus clouds, however, which were observed to be columns, rosettes, and bullet rosettes in shape, are usually several hundreds of micrometers in size, while smaller micrometer-sized particles appear to be spherical \citep{lawson1998}.

For this study, we therefore assumed small spherical particles with $p_1 = 7$ and $p_2 = \SI{0.1}{\um}$, defining the particle size distribution to apply the Mie scattering theory.
These parameters correspond to an effective particle radius of \SI{1}{\um} and an effective variance of 0.1 (see Eq.~\ref{eq:size-distiribution-parameter}).
Probing larger and irregularly shaped particles, such as water-ice particles \citep[see][]{karalidi2012b}, was beyond the scope of this paper.

The cloud layer was located between \SI{1}{bar} and \SI{0.1}{bar}.
The choice of the size parameter and pressure values was based on previously motivated assumptions of micrometer-sized particles at pressures of several hundred \si{millibar} in cloudy atmospheres, for example, in the atmosphere of Jupiter \citep{west1986} and Venus \citep{hansen1974a, knibbe1997}.
The parameter values for the standard size distribution are also in agreement with the cloud model by \citet{stam2004} for a Jupiter-like planet.
In addition, in the case of HD 189733b, three-dimensional radiative-hydrodynamics with kinetic microphysical mineral cloud formation processes show that mean cloud particle sizes are typically submicron to micron at pressures lower than \SI{1}{bar} \citep{lee2016}.

\citet{perez-hoyos2005} presented a study of the vertical structure of clouds in the atmosphere of Saturn and found a strong latitudinal dependence on its optical thickness in the visible wavelength region, ranging from 20 to 45 at the equator, but 5 at the pole.
Since the main goal of this study was to investigate the influence of different cloud compositions, and to avoid introducing too many variables, our planetary model had a homogenous cloud cover.
The vertical optical depth of the cloud layer was chosen to be equal to 10 at \SI{0.5}{\um}, which falls in the range assumed for the clouds of Saturn \citep{perez-hoyos2005}, and also represents the approximate visible mean optical depth of the clouds of Jupiter as proposed by \citet{sato1979}.

Depending on the temperature profile in the atmosphere and its composition, clouds can condensate in various pressure regions \citep[e.g.,][]{lodders2002, visscher2006, visscher2010}, with condensates grown up to several hundreds of micrometers in size \citep{helling2008, lee2015}.
However, investigating the latitudinal and longitudinal dependence on the height and optical depth of clouds as well as the dependence on the particle size and its influence on the reflected polarized flux were beyond the scope of the current study.
Therefore, the planetary model discussed above served as a reference model for all cloud types, regardless of the composition.

\begin{table}
    \centering
    \caption{Different cloud condensates, and the reference for the complex refractive index.}
    \begin{tabular}{l l}
        \hline\hline
        Formula & Reference \\
        \hline
        Al$_2$O$_3$     & \citet{koike1995} \\
        C               & \citet{draine2003} \\
        CaTiO$_3$       & \citet{ueda1998} \\
        CH$_4$          & \citet{martonchik1994} \\
        Cr              & Lynch \& Hunter in \citet{palik1991} \\
        Fe              & Lynch \& Hunter in \citet{palik1991} \\
        FeO             & \citet{henning1995}\tablefootmark{a} \\
        FeS             & \citet{pollack1994} \\
        Fe$_2$O$_3$     & A.H.M.J. Triaud, unpublished\tablefootmark{a} \\
        Fe$_2$SiO$_4$   & \citet{fabian2001}\tablefootmark{a} \\
        H$_2$O          & \citet{warren2008} \\
        KCl             & Palik in \citet{palik1985} \\
        MgAl$_2$O$_4$   & Tropf \& Thomas in \citet{palik1991} \\
        MgFeSiO$_4$     & \citet{dorschner1995}\tablefootmark{a} \\
        MgO             & Roessler \& Huffman in \citet{palik1991} \\
        MgSiO$_3$       & \citet{jaeger2003}\tablefootmark{a} \\
        Mg$_2$SiO$_4$   & \citet{jaeger2003}\tablefootmark{a} \\
        MnS             & \citet{huffman1967} \\
        NaCl            & Eldrige \& Palik in \citet{palik1985} \\
        Na$_2$S         & \citet{khachai2009} \\
        NH$_3$          & \citet{martonchik1984} \\
        SiO             & Philipp in \citet{palik1985} \\
        SiO$_2$         & Philipp in \citet{palik1985} \\
        TiO$_2$         & \citet{zeidler2011, siefke2016}\tablefootmark{a} \\
        ZnS             & Palik \& Addamiano in \citet{palik1985} \\
        \hline
    \end{tabular}
    \tablefoot{
    Most of the data is provided by \citet{kitzmann2018} as part of the
open-source \href{https://github.com/exoclime}{Exoclime Simulation Platform}.\\
    \tablefoottext{a}{\href{https://www.astro.uni-jena.de/Laboratory/OCDB/index.html}{Database of Optical Constants for Cosmic Dust}, Laboratory Astrophysics Group of the AIU Jena.}
    }
    \label{tab:cloud-condensates}
\end{table}

The compositions considered here (see Table~\ref{tab:cloud-condensates}) are expected to condensate in different types of atmospheres of extrasolar planets and brown dwarfs \citep{kitzmann2018}.
Methane (CH$_4$), ammonia (NH$_3$), or water (H$_2$O) condensate at temperatures of about \SI{80}{K} to $\sim$\SI{350}{K} \citep{carlson1988, lodders2002}.
Sulfide or chloride clouds, such as zinc sulfide (ZnS), sylvite (KCl), or sodium sulfide (Na$_2$S), condensate in the range of about \SI{700}{K} to $\sim$\SI{1100}{K} and are therefore expected in atmospheres of Y and T dwarfs \citep{morley2012}.
Iron, corundum (Al$_2$O$_3$), or silicate clouds, such as enstatite (MgSiO$_3$) or forsterite (Mg$_2$SiO$_4$), usually condensate at high temperatures ($\gtrsim$\SI{1300}{K}) in atmospheres of L dwarfs and deep in the atmospheres of T dwarfs \citep{lodders2003, visscher2010}.

Clouds can condensate in either liquid or solid phase.
In the case of iron and silicates, liquid clouds are usually found at much higher temperatures and pressures.
For example, liquid iron forms at $\geq$\SI{1809}{K}, liquid enstatite at $\geq$\SI{1851}{K}, and liquid forsterite at $\geq$\SI{2163}{K} \citep{lodders1999}.
The liquid condensates may also dissolve other elements and have complex compositions \citep{lodders1999}.
In addition, optical properties of these iron or silicates compositions exist mainly for the solid phase in the literature.

When we compare the refractive index of the solid and liquid states of CH$_4$, H$_2$O, or NH$_3$, for instance, measurements show a difference in the real part of the refractive index of about 0.05 at \SI{0.5}{\um} for CH$_4$ \citep[interpolated]{martonchik1994}, 0.03 at \SI{0.5}{\um} for H$_2$O \citep{hale1973, warren2008}, or 0.06 for the sodium D line for NH$_3$ \citep{marcoux1969}.
Although these differences in the real part of the refractive index are minimal, they have an impact on the resulting optical properties, such as cross-sections and the scattering matrix.
For H$_2$O, the reflected radiation of Earth-like exoplanets with liquid and solid water clouds was studied, for example, by \citet{karalidi2012b}.
In this study, we assumed Uranus- and Neptune-like or Jupiter-like cloud compositions, that is, methane-ice clouds \citep{lindal1987, lindal1990, lindal1992} or ammonia-ice and water-ice clouds \citep{atreya1999} are expected, respectively.
Furthermore, \citet{morley2014} modeled atmospheres of Y dwarfs and exoplanets and found that water condenses in the upper atmosphere to form ice clouds.
Thus, we investigated the case of solid cloud condensates in the atmosphere.

\section{Results}
\label{sec:results}

Our goal was to determine the feasibility of distinguishing various cloud condensates expected in exoplanetary atmospheres using polarimetric observations.
Therefore, we first reviewed the characteristic properties of the polarization resulting from single scattering by individual cloud particles.
Subsequently, we investigated the scattered radiation of a three-dimensional planetary atmosphere consisting of gas and cloud layers as described in Sect.~\ref{sec:model-setup}, taking multiple scattering events as well as absorption of the cloud particles and the ground layer into account.

\subsection{Polarization characteristics of individual particles}
\label{subsec:single-scattering}

\begin{figure*}
    \centering
    \includegraphics[width=18cm]{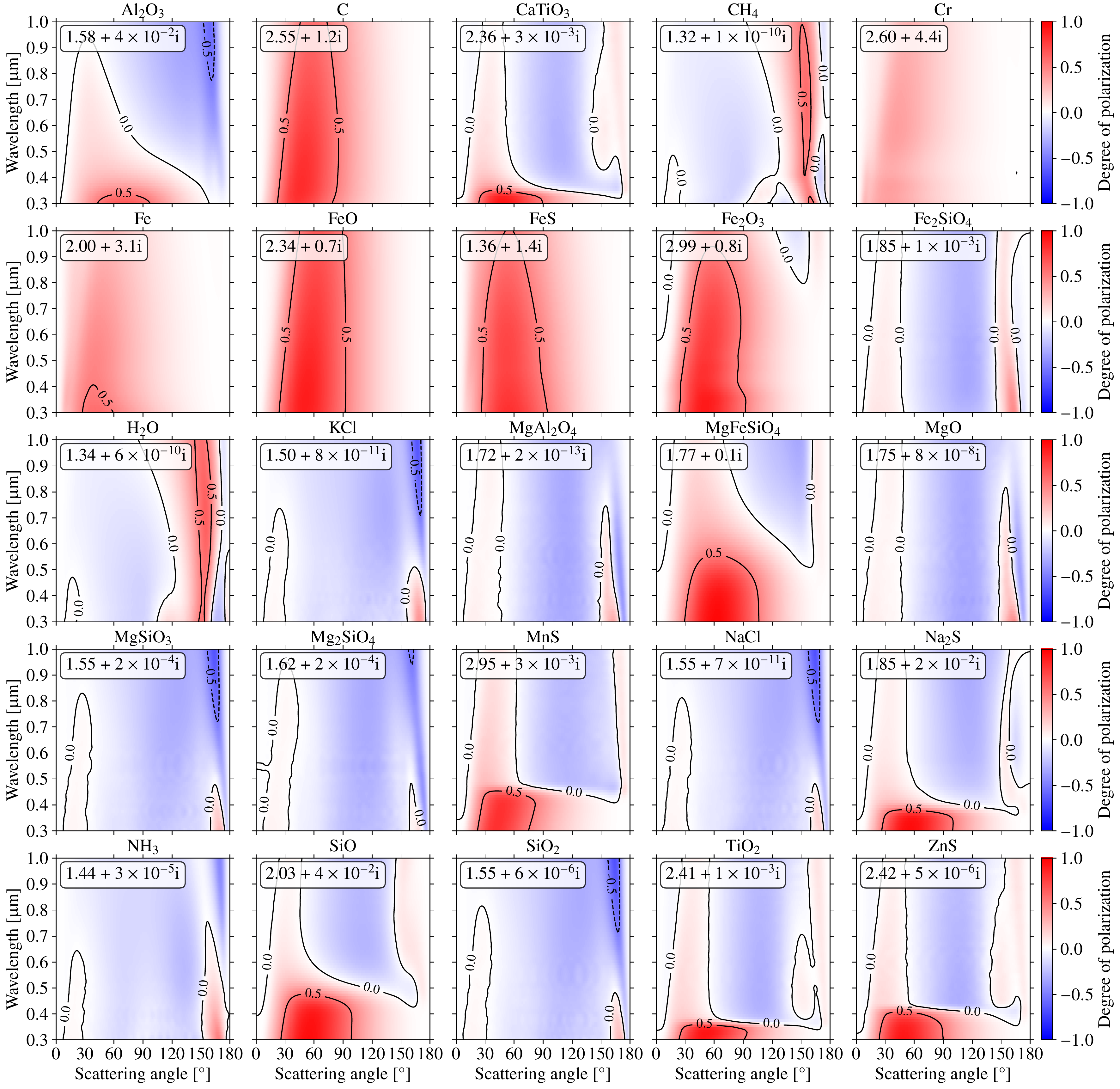}
    \caption{Degree of signed polarization $P_\mathrm{s}$ of single-scattered radiation (see Eq.~\ref{eq:single-scattering-polarization}) for various cloud types as a function of wavelength and scattering angle ($\theta = \ang{0}$: forward scattering).
    In addition, the approximate value of the complex refractive index at a wavelength of \SI{0.5}{\um} is shown in the top left corner of each figure.
    See Sect.~\ref{subsec:single-scattering} for details.}
    \label{fig:single-scattering}
\end{figure*}

As a preparatory study for the investigation of the polarization due to scattering in exoplanetary clouds, we analyzed and reviewed characteristic features that can be found in the polarization due to single scattering by potential cloud particles \citep{hansen1974b, bailey2007, karalidi2011}.
For this purpose, we calculated the corresponding scattering matrix, based on the wavelength-dependent complex refractive index of the various condensates (see Table~\ref{tab:cloud-condensates}).
Figure~\ref{fig:single-scattering} shows the degree of polarization given by Eq.~\ref{eq:single-scattering-polarization} for single-scattered radiation as a function of wavelength and scattering angle for various compositions.
In general, materials with similar refractive indices also show a similar polarization pattern.
Furthermore, materials with a high imaginary part of the refractive index ($\gtrsim$\num{e-1}) mainly show a positive polarization, while materials with a low imaginary part of the refractive index ($\lesssim$\num{e-3}) also show a negative polarization.

For H$_2$O and CH$_4$ (see Fig.~\ref{fig:single-scattering}) with a real part of the refractive index of about 1.31 and 1.32, respectively, at a wavelength of \SI{0.5}{\um}, there is a broad plateau of positive polarization at a scattering angle of approximately \ang{150}.
This is the result of rays that are reflected inside the particle once, also called the primary rainbow. This has been discussed, for example, by \citet{liou1971}, \citet{hansen1974b}, or \citet{bailey2007}.
The plateau of positive polarization ranges over the entire considered wavelength range.
For scattering angles below \ang{90}, the degree of polarization mainly has a negative sign, which is a consequence of rays that are refracted twice at the particle \citep{hansen1974b}.
At scattering angles of about \ang{20}, there is a second plateau of positive polarization.
This is a result of rays that are externally reflected once at the particle \citep{hansen1974b}.
This plateau ranges up to a wavelength of about \SI{0.5}{\um}.

For increasing wavelengths, the rainbow is shifted toward shorter wavelengths and larger scattering angles \citep[e.g.,][]{liou1971, hansen1974b, bailey2007}.
This is the case for NH$_3$ with a real part of the refractive index of about 1.44 at a wavelength of \SI{0.5}{\um}, or for KCl with a real part of the refractive index of about 1.5 at \SI{0.5}{\um}.
For increasing wavelengths, a plateau of negative polarization forms at large scattering angles of about \ang{170} due to rays that are incident on the edge of the particle and internally reflected once as well \citep{hansen1974b}.
In addition, at small scattering angles of about \ang{20}, the plateau of positive polarization ranges up to a wavelength of about \SI{0.6}{\um} for NH$_3$ and \SI{0.7}{\um} for KCl.
Similar features were discussed by \citet{hansen1974a} in the context of the analysis of scattered light polarimetry of Venus.
The real part of the refractive index of sulfuric acid clouds in the atmosphere of Venus is about 1.44 and is therefore similar to the real refractive index of ammonia.

This shift of the rainbow with increasing real part of the refractive index is valid up to a refractive index of about 1.547 \citep{nussenzveig1969}.
Hence, for compositions such as MgAl$_2$O$_4$, MgO, or Fe$_2$SiO$_4$, the positive polarization shifts back to longer wavelengths, as also discussed by \citet{hansen1974b}.
In addition, for a real part of the refractive index of $\gtrsim$1.7, the plateau of positive polarization at small scattering angles ranges over the entire considered wavelength region.
However, a peak of positive polarization at the end of this plateau, which is described by \citet{hansen1974b} and is a result of anomalous diffraction \citep{van-de-hulst1957}, is not visible here.

For a real part of the refractive index $>$2, such as CaTiO$_3$, TiO$_2$, or ZnS, two plateaus of positive polarization form at large scattering angles.
One plateau is located at a scattering angle of about \ang{150} and ranges from a wavelength of \SI{0.4}{\um} to $\sim$\SI{0.7}{\um}.
The second plateau is located at scattering angles of about \ang{180} and ranges over the entire considered wavelength region.

For an increasing imaginary part of the refractive index, the amount of radiation traveling through the particle decreases.
As a consequence, features such as negative polarization due to double refraction, or positive polarization due to internal reflections are lost, and radiation reflected from outside the particle with a positive degree of polarization dominates the polarization pattern \citep{hansen1974b}.
Thereby, the maximum degree of polarization is shifted toward $\ang{180} - 2\theta_\mathrm{B}$, where $\theta_\mathrm{B}$ is the Brewster angle \citep{brewster1815}.
This is the case for C, Cr, Fe, FeO, FeS, or Fe$_2$O$_3$.
Furthermore, for compositions with a very large imaginary part of the refractive index ($\gtrsim$3), such as Fe or Cr, the degree of linear polarization of single-scattered radiation decreases compared to compositions such as FeO or FeS with an imaginary part of the refractive index of $\lesssim$1.4 at \SI{0.5}{\um}.
This applies to compositions with a high absolute value of the complex refractive index in general.

For compositions such as Al$_2$O$_3$, MgFeSiO$_4$, MnS, or SiO, there is a high imaginary part of the refractive index at short wavelengths and a small imaginary part of the refractive index at long wavelengths.
At short wavelengths, the scattered radiation is therefore dominated by a high degree of positive polarization, while with increasing wavelength, the polarization decreases and becomes negative at a characteristic wavelength.

\subsection{Polarization of an exoplanetary atmosphere}
\label{subsec:exoplanet-scattering}

\begin{figure*}
    \centering
    \includegraphics[width=18cm]{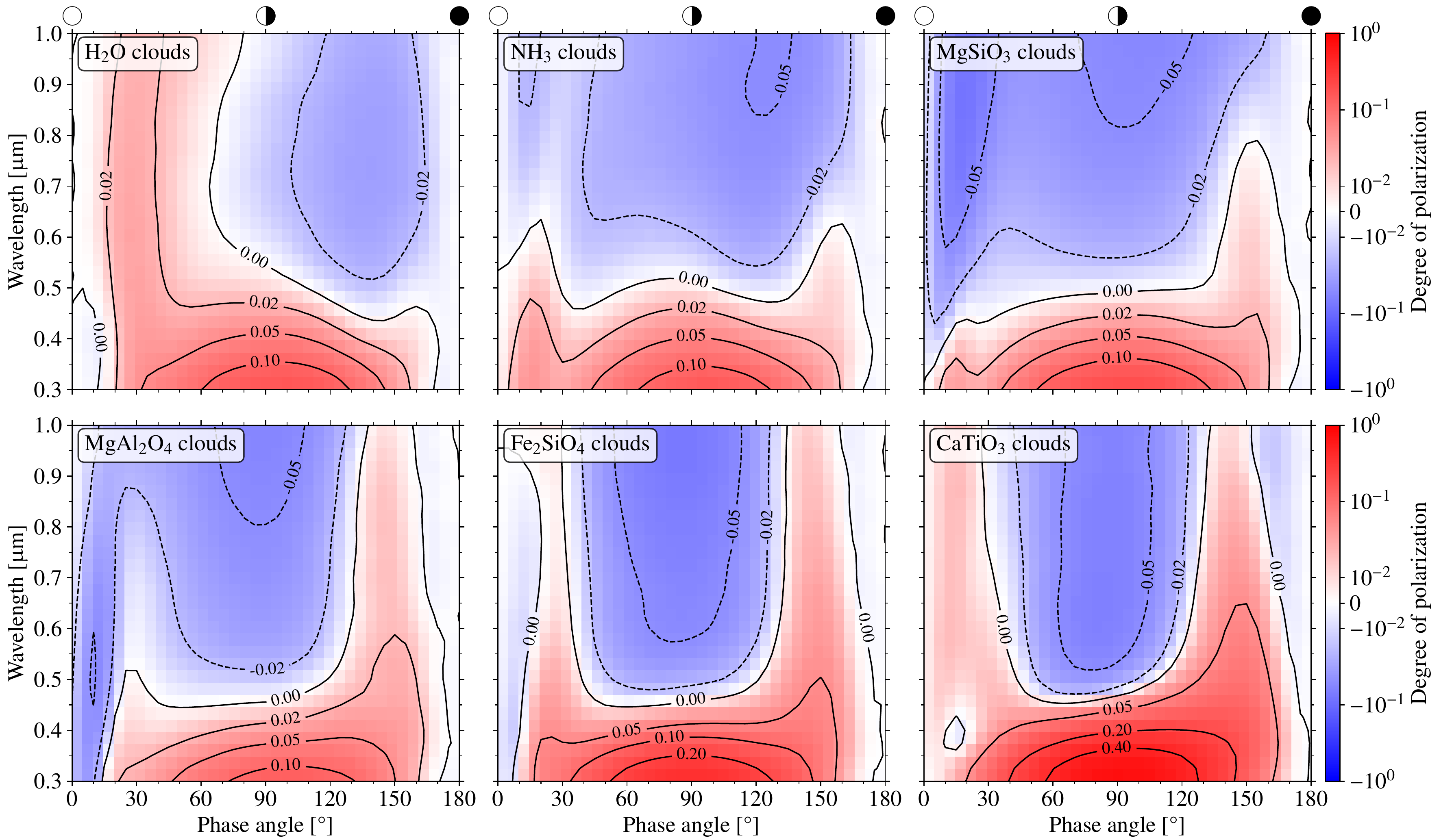}
    \caption{Degree of signed polarization $P_\mathrm{s}$ (see Eq.~\ref{eq:signed-linear-polarization}) as a function of wavelength and phase angle for various cloud compositions.
    As indicated by the spherical symbols at the top, a phase angle of \ang{0} corresponds to the planet in full phase.
    See Sect.~\ref{subsubsec:var-real} for details.}
    \label{fig:exoplanet-polarization-1}
\end{figure*}

\begin{figure*}
    \centering
    \includegraphics[width=18cm]{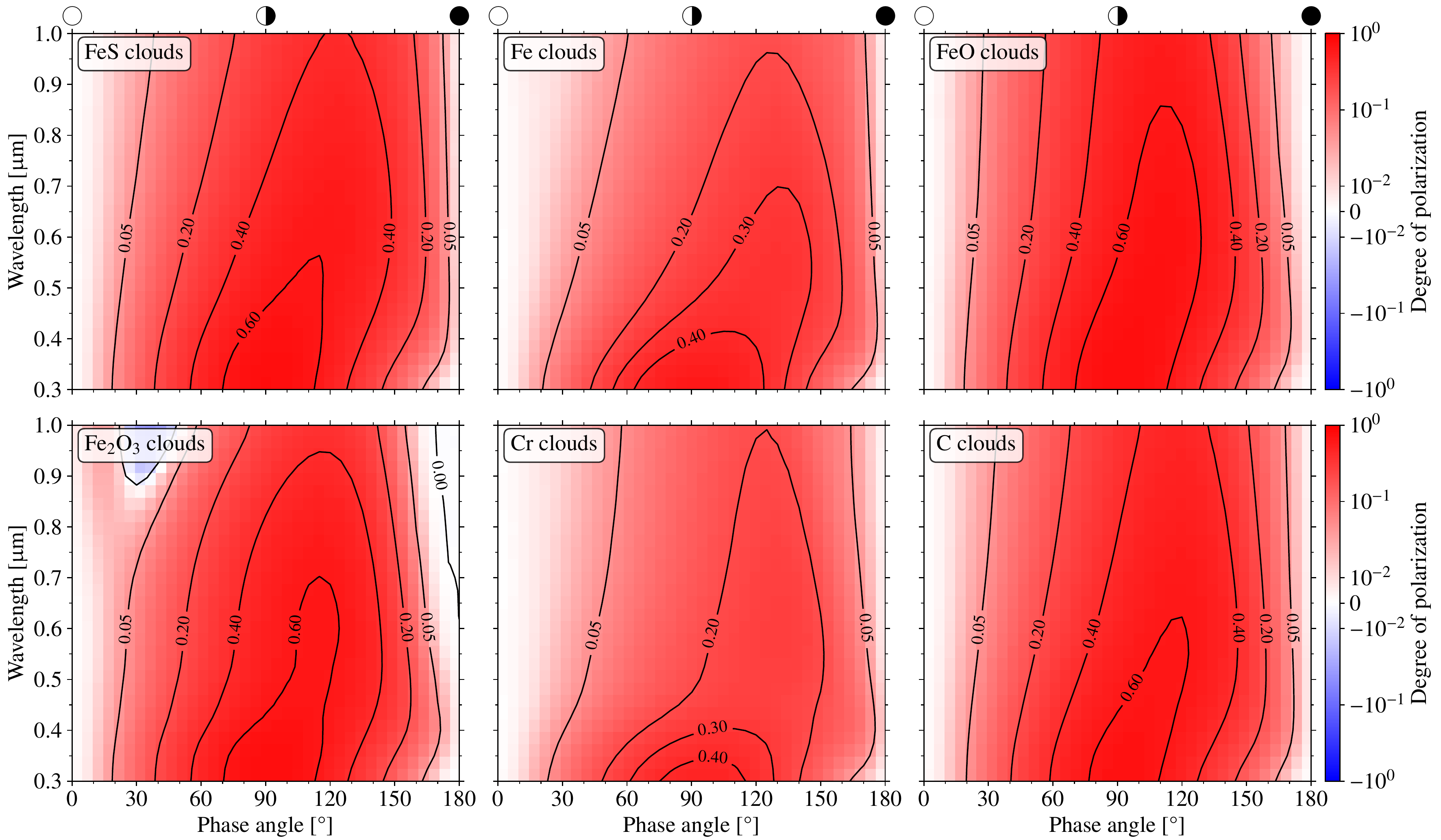}
    \caption{Same as Fig.~\ref{fig:exoplanet-polarization-1}, but for different cloud compositions.
    See Sect.~\ref{subsubsec:large-imag} for details.}
    \label{fig:exoplanet-polarization-2}
\end{figure*}

Although it is possible to characterize the refractive index of the material by its single-scattered polarized radiation, as discussed in Sect.~\ref{subsec:single-scattering}, multiple scattering and additional Rayleigh scattering by gaseous particles influence the polarized radiation scattered in a cloudy exoplanetary atmosphere \citep[e.g.,][]{karalidi2011}.
It therefore remained to be investigated whether the considered cloud compositions with their characteristic polarization features identified in Sect.~\ref{subsec:single-scattering} due to, for instance, a wavelength-dependent imaginary part of the refractive index, are distinguishable in this significantly more complex environment.
Based on the model described in Sect.~\ref{sec:model-setup}, we calculated the scattered polarized radiation of a cloudy planetary atmosphere.
In Sects.~\ref{subsubsec:var-real} to \ref{subsubsec:sim-real}, we investigated and compared different cloud compositions depending on the properties of their refractive index.

At shorter wavelengths ($\sim$\SI{0.3}{\um}), the degree of polarization for all atmospheric models is in general positive, with a maximum value around a phase angle of \ang{90} due to Rayleigh scattering of the gaseous layers above the clouds, as discussed, for example, by \citet{hansen1974a}, \citet{stam2008}, or \citet{karalidi2011}.
However, with increasing wavelength, the impact of the gaseous particles weakens because of the strong wavelength dependence on the Rayleigh scattering cross-section (${\propto} \lambda^{-4}$).
Therefore, the degree of polarization of the cloud particles starts to dominate the net scattered radiation \citep{karalidi2011}.
As a result of multiple scattering in the atmosphere, the degree of polarization is lower than that of single scattering \citep{hansen1974b, karalidi2011}.
However, although multiple scattering decreases the degree of polarization, it does not change the sign of polarization because the reflected radiation is dominated by single-scattered radiation from the upper layers of the clouds \citep{karalidi2011}.
Thus, the features discussed in Sect.~\ref{subsec:single-scattering} are still imprinted on the scattered polarized flux.
In the following and in contrast to Fig.~\ref{fig:single-scattering}, we describe the scattered polarized flux as a function of the planetary phase angle $\alpha$.
For single-scattered radiation, it is $\alpha = \ang{180} - \theta$.

\subsubsection{Clouds with various real parts of the refractive index}
\label{subsubsec:var-real}

In Fig.~\ref{fig:exoplanet-polarization-1}, the signed degree of linear polarization ($P_\mathrm{s}$) of various model atmospheres is shown as a function of wavelength and phase angle.
The model atmospheres have clouds consisting of H$_2$O, NH$_3$, MgSiO$_3$, MgAl$_2$O$_4$, Fe$_2$SiO$_4$, or CaTiO$_3$.
These different cloud compositions have a small imaginary part of the refractive index ($\lesssim$\num{e-3} at \SI{0.5}{\um}) in common and cover a broad range of the real part of the refractive index ($\sim$1.34 to $\sim$2.36 at \SI{0.5}{\um}).
In this case, our simulations confirm that the characteristic polarization features due to the real part of the refractive index can be used to distinguish these different materials and to characterize the refractive index.
Especially the locations of the positive degree of polarization at small phase angles (or at large scattering angles) have been discussed previously by various authors \citep[e.g.,][]{hansen1974a, bailey2007, karalidi2011}.

For clouds with a real part of the refractive index of about 1.55 to 1.75 such as MgSiO$_3$ and MgAl$_2$O$_4$, the plateau of negative polarization at small phase angles ($\lesssim$\ang{10}) is visible for the entire considered wavelength region.
In addition, for cloud condensates with a real part of the refractive index of $\gtrsim$1.75 such as MgAl$_2$O$_4$, Fe$_2$SiO$_4$, and CaTiO$_3$, the plateau of the positive degree of polarization at large phase angles ranges over the entire considered wavelength region as well.
For CaTiO$_3$ clouds, there is a small spot of zero polarization at a phase angle of about \ang{15} and at a wavelength of about \SI{0.38}{\um}, surrounded by a positive degree of polarization.
Furthermore, the degree of polarization increases at small wavelengths ($\sim$\SI{0.3}{\um}) for CaTiO$_3$ because the imaginary part of the refractive index increases slightly at these wavelengths.

Although \citet{karalidi2011} considered liquid-water clouds, the results of our simulations for water clouds are in agreement with those obtained by \citet{karalidi2011} because the authors used similar parameter values for the size distribution and the difference in the refractive index comparing liquid and solid water is small (see Sect.~\ref{sec:model-setup}).
In contrast, in the study by \citet{karalidi2012b}, the plateau of positive polarization at phase angles about \ang{30} solely arises from internal reflections of liquid-water clouds and not from water-ice particles because the authors assumed nonspherical ice crystals.

The polarization curves of close-in extrasolar giant planets by \citet{seager2000} or the polarization curves applied to HD 189733b by \citet{kopparla2016} and \citet{bailey2018} assumed cloud condensates, such as iron or silicates.
\citet{seager2000} considered a homogeneous MgSiO$_3$-Al$_2$O$_3$-Fe cloud, and \citet{kopparla2016} assumed a silicate cloud with a refractive index of $1.68 + 10^{-4}\mathrm{i}$.
The authors of these two studies observed a peak of polarization at small phase angles, which agrees with our peak of negative polarization at small phase angles.
\citet{bailey2018} compared clouds composed of MgSiO$_3$, Mg$_2$SiO$_4$, Al$_2$O$_3$, or Fe.
However, the authors considered particle sizes of $r_\mathrm{eff} \leq \SI{0.1}{\um}$, which results in size parameters of $\lesssim$1.4. The scattering is therefore almost Rayleigh scattering.
Thus, the polarization shows a maximum centered around a phase angle of \ang{90}.

\subsubsection{Clouds with a large imaginary part of the refractive index}
\label{subsubsec:large-imag}

In Fig.~\ref{fig:exoplanet-polarization-2}, the signed degree of linear polarization ($P_\mathrm{s}$) is shown for model atmospheres with clouds consisting of FeS, Fe, FeO, Fe$_2$O$_3$, Cr, or C.
These different cloud compositions cover a broad range of the real part of the refractive index ($\sim$1.36 to $\sim$2.99 at \SI{0.5}{\um}) and have a large imaginary part of the refractive index ($\gtrsim$0.7 at \SI{0.5}{\um}) in common.
Due to the high imaginary part of the refractive index, the single-scattering albedo of the cloud condensates decreases, resulting in a lower impact of multiple scattering and, thus, an increase in the degree of polarization.
However, for Fe and Cr clouds, the polarization decreases because the degree of polarization after single scattering decreases (see Sect.~\ref{subsec:single-scattering}).
The maximum value of the degree of polarization is $\sim$0.59 and $\sim$0.47 for Fe and Cr clouds, respectively.
For atmospheres composed of FeS, FeO and Fe$_2$O$_3$ clouds, the maximum value of the degree of polarization is $\sim$0.75.
In contrast to purely Rayleigh scattering, the maximum degree of polarization is shifted to larger phase angles for these compositions (see Sect.~\ref{subsec:single-scattering}).
At long wavelengths, the maximum degree of polarization is shifted to phase angles of about \ang{125} for FeS, Fe, and Cr clouds, while for Fe$_2$O$_3$ and C clouds, the maximum is shifted to phase angles of about \ang{115}, and to about \ang{110} for FeO clouds.
Since the imaginary part of the refractive index decreases at longer wavelengths for Fe$_2$O$_3$, there is a plateau of negative polarization for atmospheres composed of Fe$_2$O$_3$ at phase angles of about \ang{30}.

\subsubsection{Clouds with a variable imaginary part of the refractive index}
\label{subsubsec:var-imag}

\begin{figure*}
    \centering
    \includegraphics[width=18cm]{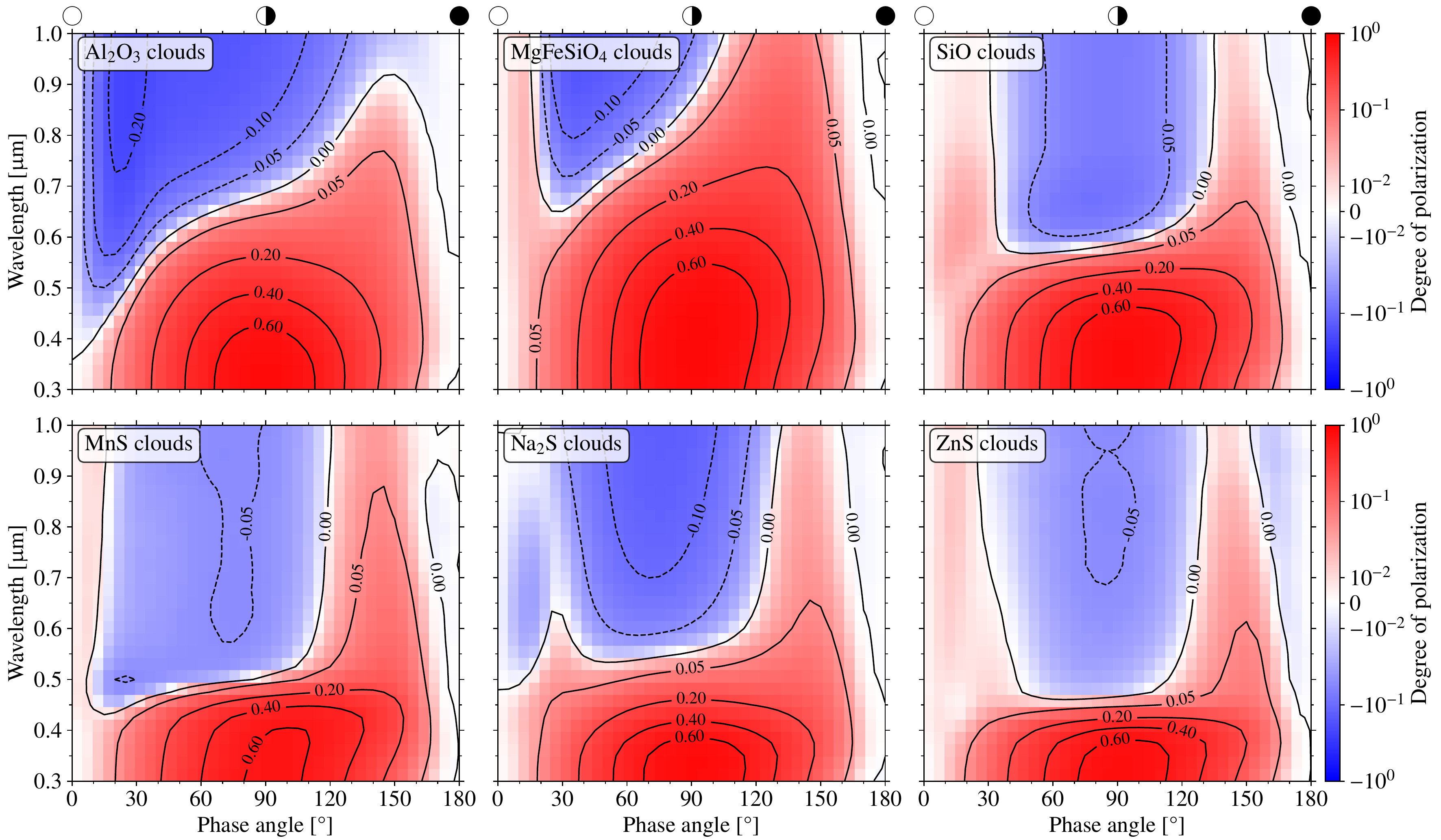}
    \caption{Same as Fig.~\ref{fig:exoplanet-polarization-1}, but for different cloud compositions.
    See Sect.~\ref{subsubsec:var-imag} for details.}
    \label{fig:exoplanet-polarization-3}
\end{figure*}

\begin{figure*}
    \centering
    \includegraphics[width=18cm]{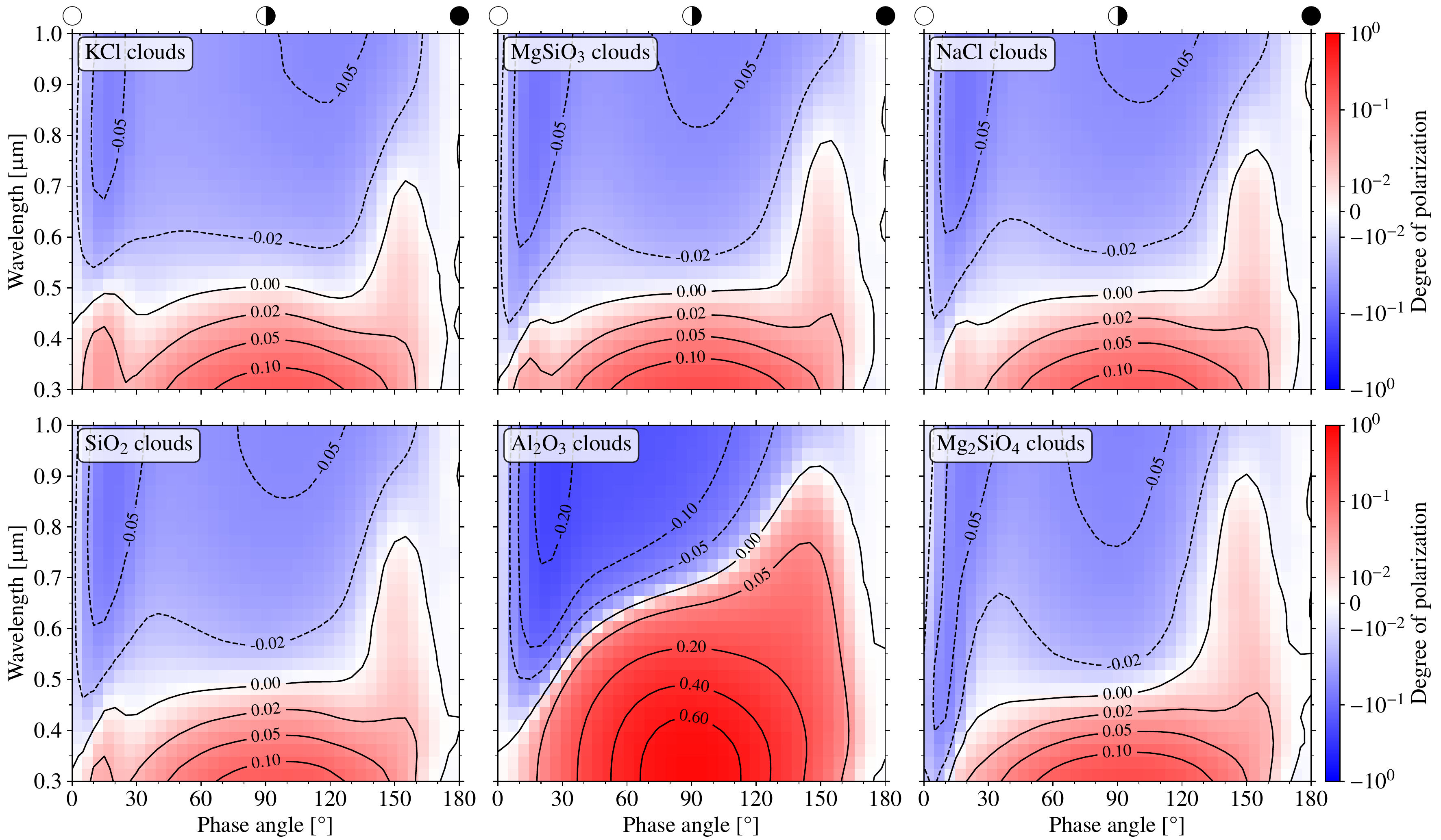}
    \caption{Same as Fig.~\ref{fig:exoplanet-polarization-1}, but for different cloud compositions.
    See Sect.~\ref{subsubsec:sim-real} for details.}
    \label{fig:exoplanet-polarization-4}
\end{figure*}

In Fig.~\ref{fig:exoplanet-polarization-3}, the signed degree of linear polarization ($P_\mathrm{s}$) is shown for model atmospheres with clouds consisting of Al$_2$O$_3$, MgFeSiO$_4$, SiO, MnS, Na$_2$S, or ZnS.
These different cloud compositions cover a broad range of the real part of the refractive index ($\sim$1.58 to $\sim$2.95 at \SI{0.5}{\um}) and have a varying imaginary part of the refractive index in the considered wavelength region.
For example, for MnS, the imaginary part of the refractive index decreases from about 1.3 at \SI{0.3}{\um} to about \num{2.7e-5} at \SI{1}{\um}.
In this case, the reflected radiation has solely a high degree of positive polarization at short wavelengths, but characteristic positive as well as negative polarization features at long wavelengths.
Thus, at wavelengths $\gtrsim$\SI{0.6}{\um}, the plateaus of positive and negative polarization determine the real part of the refractive index of the composition.
At short wavelengths, the behavior of the positive polarization is linked to the imaginary part of the refractive index.
In contrast to Al$_2$O$_3$ and MgFeSiO$_4$ clouds, the positive polarization for an atmosphere composed of SiO, MnS, Na$_2$S, and ZnS clouds decreases rapidly with increasing wavelength.
At a characteristic wavelength of about \SI{0.5}{\um} to \SI{0.6}{\um}, depending on the material, the sign of the polarization changes.
This is the result of the strong decrease of the imaginary part of the refractive index at these wavelengths.
For atmospheres composed of Al$_2$O$_3$ and MgFeSiO$_4$ clouds, no such characteristic wavelength can be defined.
Instead, the sign of polarization changes at different wavelengths for different planetary phase angles.
In addition, for Al$_2$O$_3$, MgFeSiO$_4$, and Na$_2$S clouds, the degree of negative polarization at long wavelengths is larger than for cloud compositions such as SiO, MnS, or ZnS.
This is due to the larger imaginary part of the refractive index of Al$_2$O$_3$, MgFeSiO$_4$, and Na$_2$S at these wavelengths, resulting in a smaller single-scattering albedo and, thus, a lower impact of multiple scattering.

\subsubsection{Clouds with an equal real part of the refractive index}
\label{subsubsec:sim-real}

In Fig.~\ref{fig:exoplanet-polarization-4}, the signed degree of linear polarization ($P_\mathrm{s}$) is shown for model atmospheres with clouds consisting of KCl, MgSiO$_3$, NaCl, SiO$_2$, Al$_2$O$_3$, and Mg$_2$SiO$_4$, which have real part of the refractive index of about 1.50 to 1.62 at \SI{0.5}{\um}.
Except for Al$_2$O$_3$, the compositions have a small imaginary part of the refractive index ($\lesssim$\num{e-4} at \SI{0.5}{\um}) as well.
The plateau of positive polarization at large phase angles and the plateau of negative polarization at small phase angles show similar characteristics, regardless of the material.
The most significant difference is found for an atmosphere consisting of Al$_2$O$_3$ clouds.
The plateau of positive polarization at large phase angles and at long wavelengths reveals a real part of the refractive index, similar to Mg$_2$SiO$_4$ clouds.
However, the higher degree of polarization for Al$_2$O$_3$ clouds compared to the other compositions indicates a smaller single-scattering albedo and, thus, a larger imaginary part of the refractive index.

Compositions with a similar behavior of real as well as imaginary part of the refractive index in the considered wavelength region, such as MgSiO$_3$, NaCl, or SiO$_2$, cannot be distinguished from each other.
However, NaCl and MgSiO$_3$ have different condensation temperatures of $\sim$\SI{800}{K} \citep{burrows1999} and $\sim$\SI{1600}{K} \citep{visscher2010}, respectively.
Thus, NaCl potentially condensates in atmospheres of T dwarfs, for example, while MgSiO$_3$ will more likely be present in atmospheres of M dwarfs or deep in atmospheres of L dwarfs.
In addition, as discussed by \citet{visscher2010}, if MgSiO$_3$ condensates in the atmosphere, it removes silicon from the gas phase, preventing SiO$_2$ ($\sim$\SI{1550}{K}, \citealt{visscher2010}) from condensating.

\section{Discussion}
\label{sec:discussion}

\begin{figure*}
    \centering
    \includegraphics[width=18cm]{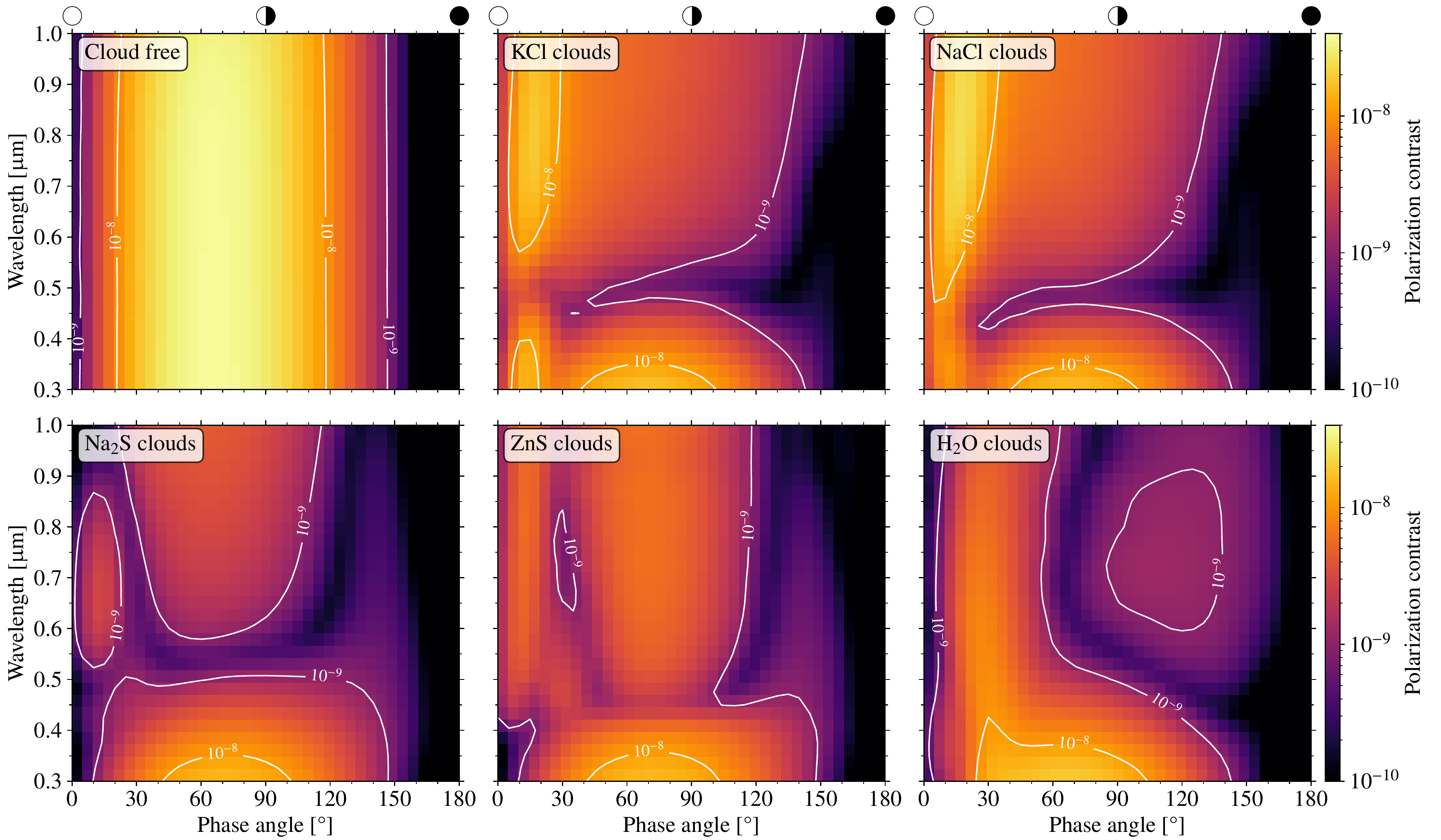}
    \caption{Polarization contrast $C_\mathrm{pol}$ (see Eq.~\ref{eq:polarization-contrast}) as a function of wavelength and phase angle for various cloud compositions and a cloud-free atmosphere (top left).
    As indicated by the spherical symbols at the top, a phase angle of \ang{0} corresponds to the planet in full phase.
    See Sect.~\ref{sec:discussion} for details.}
    \label{fig:polarization-contrast}
\end{figure*}

In this section, we address selected parameters and corresponding questions that also have an impact on the net polarization, but were not covered in our simulations.
The optical depth of the cloud layer in this study had a fixed value of 10 at \SI{0.5}{\um}.
An increasing optical depth of the cloud layer affects the degree of net polarization because the resulting increasing multiple scattering decreases the net degree of polarization \citep{karalidi2011, bailey2018}.
On the other hand, a decreasing cloud optical depth would cause the Rayleigh scattering to dominate the net degree of polarization.
This is similar to the case of a cloud layer that is located in higher pressure regions because the optical depth of gaseous particles increases above the cloud layer.
Information about the polarization of deep cloud layers is lost at optical depths $\gtrsim$2 of the upper layers \citep{buenzli2009}.
For a detailed discussion of cloud top pressure and cloud optical depth, see \citet{karalidi2011}.

The size distribution of the cloud particles is another crucial parameter determining the scattering properties of the particles.
While the effective radius causes the degree of polarization to shift along the wavelength axis, increasing the effective variance causes various polarization features to smooth out \citep{hansen1974a, hansen1974b}.
For smaller particles for a given observing wavelength, that is, smaller size parameters, the scattering is in the Rayleigh regime, resulting in a loss of the characteristic features in the polarization.
For a detailed study of various particle sizes, see, for instance, \citet{hansen1974a}, \citet{hansen1974b}, \citet{seager2000}, \citet{karalidi2011}, or \citet{karalidi2012b}.

In our study, the planetary atmosphere had a homogeneous cloud coverage that covered the entire planet.
However, the cloud coverage fraction and an asymmetric coverage of hemispheres have an impact on the scattered flux as well.
The influence of the cloud coverage was studied in detail by \citet{karalidi2012a}, and \citet{karalidi2013}, and the retrieval of the cloud coverage of Earth-like exoplanets was studied, for example, by \citet{rossi2017}.
In addition, horizontally inhomogeneous cloud coverage also causes a nonzero circular polarization degree, which was studied by \citet{rossi2018a}.

As already mentioned in Sect.~\ref{sec:model-setup}, we ignored absorption by gaseous particles.
In general, absorption will decrease the effect of multiple scattering, thus, the degree of polarization increases.
However, the net degree of polarization can decrease as well if there are, for instance, reflecting clouds in the lower layers.
For a detailed discussion and radiative transfer computation with absorption by gaseous particles, see, for example, \citet{stam2004} or \citet{buenzli2009}.

For the degree of polarization shown in Figs.~\ref{fig:exoplanet-polarization-1}--\ref{fig:exoplanet-polarization-4}, we assumed that the exoplanet and its host star are spatially separated observationally, thus, the high fraction of unpolarized stellar radiation is not included in the net polarization.
Considering an exoplanet-star system where both components cannot be separated observationally, the ratio of the polarized planetary flux and the total stellar flux, that is, the polarization contrast, is given by \citep{hunziker2020}
\begin{equation}
    \label{eq:polarization-contrast}
    C_\mathrm{pol} = P(\alpha, \lambda) I(\alpha, \lambda) \frac{r^2}{d^2}.
\end{equation}
Here, $P$ and $I$ are the phase angle and wavelength-dependent degree of polarization and reflectivity of the planet, respectively.
At a phase angle of \ang{0}, the reflectivity is equivalent to the planetary geometric albedo.
Furthermore, $r$ and $d$ are the planetary radius and distance between planet and star, respectively.
To calculate the planetary temperature profile, we followed \citet{guillot2010} and took parameter values based on HD~209458b ($\kappa_\mathrm{th} = \SI{e-2}{cm^2.g^{-1}}$, $\kappa_\mathrm{v} = \SI{4e-3}{cm^2.g^{-1}}$, $T_\mathrm{int} = \SI{500}{K}$, \citealt{guillot2010}; $r = \SI{1.45}{R_J}$, $T_\star = \SI{6092}{K}$, \citealt{boyajian2015}; $m = \SI{0.64}{M_J}$, \citealt{snellen2010}).
However, we set $d = \SI{1}{AU}$ with an observational distance of \SI{5}{pc} to be suitable for SPHERE/ZIMPOL \citep{hunziker2020}.
Thus, we obtained temperatures of about \SI{500}{K} to $\sim$\SI{1000}{K} in the pressure region of about \SI{10}{mbar} to $\sim$\SI{1}{bar}, which means that compositions such as KCl, NaCl, Na$_2$S, or ZnS potentially form cloud condensates.

Figure~\ref{fig:polarization-contrast} shows the polarization contrast (Eq.~\ref{eq:polarization-contrast}) as a function of wavelength and planetary phase angle for various compositions that are expected to condensate in the atmosphere, and a cloud-free atmosphere for comparison.
For the cloud-free atmosphere, the polarization contrast is highest with maximum values of about \num{3.7e-8} at phase angles of about \ang{65} across the considered wavelength region.
For the cloudy atmospheres, the polarization contrast is highest at short wavelengths ($\sim$\SI{0.35}{\um}) with values of $\sim$\num{2e-8} due to Rayleigh scattering.
In addition, for NaCl clouds, the polarization contrast reaches values of $\gtrsim$\num{2e-8} at small phase angles and long wavelengths, making the plateau of negative polarization detectable, for instance, with SPHERE/ZIMPOL \citep{hunziker2020}.
For Na$_2$S and ZnS clouds, the high degree of polarization at short wavelength with a rapidly decreasing degree of polarization for increasing wavelength, as discussed in Sect.~\ref{subsubsec:var-imag}, produces a polarization contrast of $\lesssim$\num{e8} because the total reflected flux decreases with decreasing single-scattering albedo as well.

\section{Conclusions}
\label{sec:conclusions}

In the first part of this study (Sect.~\ref{subsec:single-scattering}), we investigated the single-scattered polarized radiation of various cloud condensates, which are expected in planetary atmospheres.
We reviewed characteristic properties of the polarization state depending on the scattering angle, wavelength, and refractive index of the cloud particle.
As demonstrated in selected earlier studies, we confirmed the significant impact of the chemical composition and, thus, the complex refractive index of the cloud particles on the state of polarization of the scattered radiation.

If the imaginary part of the refractive index is small, refracted rays dominate the polarization pattern \citep{hansen1974b}.
Characteristic features, such as the rainbow, depend on the real part of the refractive index \citep[e.g.,][]{liou1971, hansen1974b, bailey2007}.
However, most of the considered cloud condensates have a nonnegligible imaginary part.
For an increasing imaginary part of the refractive index, polarization features resulting from rays that are refracted or reflected inside the particle are lost, and radiation that is reflected externally dominates \citep{hansen1974b}.
For compositions with a large imaginary part of the refractive index over the entire considered wavelength region, the polarization is positive, with the maximum value shifted to smaller scattering angles \citep[see][]{hansen1974b}.
Compositions such as Al$_2$O$_3$, MgFeSiO$_4$, SiO, MnS, Na$_2$S, and SiO have a strongly varying imaginary part of the refractive index. At short wavelengths, the polarization is therefore mainly positive and becomes negative at longer wavelengths.
In between, there is a change of sign of the polarization due to the decrease of the imaginary part.
The wavelength at which the sign changes depends on the chemical composition and, thus, allows for the identification of the chemical composition of the respective condensate.

The goal of the second part of this study (Sect.~\ref{subsec:exoplanet-scattering}) was to evaluate the feasibility to distinguish between the different considered cloud materials in a planetary atmosphere.
For this purpose, we simulated the reflected polarized flux of cloudy planetary atmospheres with selected cloud compositions investigated in Sect.~\ref{subsec:single-scattering}.
These simulations were performed with the three-dimensional Monte Carlo radiative transfer code POLARIS, which was optimized to handle the radiative transfer in planetary atmospheres.
The degree of linear polarization was studied at planetary phase angles ranging from \ang{0} to \ang{180} at wavelengths from \SI{0.3}{\um} to \SI{1}{\um}.

Building on the results of the single-scattering polarization obtained in the first part of our study, we first presented results for compositions with a small imaginary, but various real parts of the refractive index in Sect.~\ref{subsubsec:var-real}.
In agreement with earlier studies \citep{hansen1974a}, our simulations confirmed that the polarization features such as the rainbow can be used to determine the refractive index of the cloud particles.

For high values of the imaginary part of the refractive index, considered in Sect.~\ref{subsubsec:large-imag}, the polarization is mainly positive and the overall degree of polarization increases.
However, each of the considered cloud composition shows a characteristic behavior of the polarization, depending on the individual imaginary part of the refractive index.
\begin{itemize}
    
    \item Cloud compositions with an imaginary part of $\gtrsim$3, such as Fe and Cr, show a decrease in the degree of polarization compared to compositions such as FeS and FeO.
    
    \item For FeS, Fe, and Cr clouds, the maximum degree of polarization at long wavelengths is shifted to phase angles of about \ang{125}, while for FeO clouds, the maximum degree of polarization is shifted to phase angles of about \ang{110}.
    
    \item For Fe$_2$O$_3$, there is a plateau of negative polarization at wavelengths of $\gtrsim$\SI{0.9}{\um} because the imaginary part of the refractive index is decreasing rapidly with increasing wavelength.
    
\end{itemize}

If the imaginary part of the refractive index strongly varies in the considered wavelength region (Sect.~\ref{subsubsec:var-imag}), the behavior of the degree of polarization and a change of the sign of the polarization degree can be linked to the behavior of the imaginary part of the refractive index of the condensate, allowing one to characterize the cloud composition.
\begin{itemize}
    
    \item Clouds composed of Al$_2$O$_3$ and MgFeSiO$_4$ show a smooth decrease of positive polarization with increasing wavelength.
    
    \item Clouds composed of SiO, MnS, Na$_2$S, and ZnS show a rapidly decreasing degree of polarization with increasing wavelength.
    In addition, at a characteristic wavelength of about \SI{0.5}{\um} to \SI{0.6}{\um}, depending on the imaginary part of the refractive index, the sign of the polarization changes.
    
\end{itemize}

Finally, for condensates with similar values of the real part of the refractive index (Sect.~\ref{subsubsec:sim-real}), a distinction of various compositions is almost impossible.
\begin{itemize}
    
    \item The most significant difference in the polarization was identified in the case of Al$_2$O$_3$ clouds.
    Here, the underlying imaginary part of the refractive index results in a high degree of positive and negative polarization at short and long wavelengths, respectively.
    
    \item However, chlorides or sulfides with similar refractive indices as silicates have a different condensation temperature.
    Thus, compositions with a similar net polarization across the considered wavelength and phase angle region, but different condensation temperatures, can be distinguished indirectly by comparison of the temperatures of the planetary atmosphere and, thus, its potential to host certain condensates/clouds.
    
\end{itemize}

In summary, most of the considered compositions that are expected to condensate in exoplanetary atmospheres, such as chlorides, sulfides, or silicates, can be distinguished via polarization measurements due to their unique wavelength-dependent real as well as imaginary part of the refractive index and/or different condensation temperatures.

\begin{acknowledgements}
    We wish to thank the anonymous referee, for suggestions improving the presentation of the results of this study.
    This research made use of
    \href{https://www.astropy.org}{Astropy}, a community-developed core Python package for Astronomy \citep{astropy2013, astropy2018},
    \href{https://matplotlib.org}{Matplotlib} \citep{hunter2007},
    \href{https://numpy.org}{Numpy} \citep{harris2020},
    \href{https://ui.adsabs.harvard.edu}{NASA's Astrophysics Data System},
    and a modified A\&A bibliography style file with clickable link in the bibliography.\footnote{\href{https://github.com/yangcht/AA-bibstyle-with-hyperlink}{\texttt{github.com/yangcht/AA-bibstyle-with-hyperlink}}}
\end{acknowledgements}

\bibliographystyle{aa}
\bibliography{bibliography.bib}

\begin{thebibliography}{126}
\expandafter\ifx\csname natexlab\endcsname\relax\def\natexlab#1{#1}\fi

\bibitem[{{Ackerman} \& {Marley}(2001)}]{ackerman2001}
{Ackerman}, A.~S. \& {Marley}, M.~S. 2001,
  \href{https://doi.org/10.1086/321540}{\apj},
  \href{https://ui.adsabs.harvard.edu/abs/2001ApJ...556..872A}{556, 872}

\bibitem[{{Astropy Collaboration} {et~al.}(2018){Astropy Collaboration},
  {Price-Whelan}, {Sip{\H{o}}cz}, {G{\"u}nther}, {Lim}, {Crawford}, {Conseil},
  {Shupe}, {Craig}, {Dencheva}, {Ginsburg}, {VanderPlas}, {Bradley},
  {P{\'e}rez-Su{\'a}rez}, {de Val-Borro}, {Aldcroft}, {Cruz}, {Robitaille},
  {Tollerud}, {Ardelean}, {Babej}, {Bach}, {Bachetti}, {Bakanov}, {Bamford},
  {Barentsen}, {Barmby}, {Baumbach}, {Berry}, {Biscani}, {Boquien}, {Bostroem},
  {Bouma}, {Brammer}, {Bray}, {Breytenbach}, {Buddelmeijer}, {Burke},
  {Calderone}, {Cano Rodr{\'\i}guez}, {Cara}, {Cardoso}, {Cheedella}, {Copin},
  {Corrales}, {Crichton}, {D'Avella}, {Deil}, {Depagne}, {Dietrich}, {Donath},
  {Droettboom}, {Earl}, {Erben}, {Fabbro}, {Ferreira}, {Finethy}, {Fox},
  {Garrison}, {Gibbons}, {Goldstein}, {Gommers}, {Greco}, {Greenfield},
  {Groener}, {Grollier}, {Hagen}, {Hirst}, {Homeier}, {Horton}, {Hosseinzadeh},
  {Hu}, {Hunkeler}, {Ivezi{\'c}}, {Jain}, {Jenness}, {Kanarek}, {Kendrew},
  {Kern}, {Kerzendorf}, {Khvalko}, {King}, {Kirkby}, {Kulkarni}, {Kumar},
  {Lee}, {Lenz}, {Littlefair}, {Ma}, {Macleod}, {Mastropietro}, {McCully},
  {Montagnac}, {Morris}, {Mueller}, {Mumford}, {Muna}, {Murphy}, {Nelson},
  {Nguyen}, {Ninan}, {N{\"o}the}, {Ogaz}, {Oh}, {Parejko}, {Parley}, {Pascual},
  {Patil}, {Patil}, {Plunkett}, {Prochaska}, {Rastogi}, {Reddy Janga},
  {Sabater}, {Sakurikar}, {Seifert}, {Sherbert}, {Sherwood-Taylor}, {Shih},
  {Sick}, {Silbiger}, {Singanamalla}, {Singer}, {Sladen}, {Sooley},
  {Sornarajah}, {Streicher}, {Teuben}, {Thomas}, {Tremblay}, {Turner},
  {Terr{\'o}n}, {van Kerkwijk}, {de la Vega}, {Watkins}, {Weaver}, {Whitmore},
  {Woillez}, {Zabalza}, \& {Astropy Contributors}}]{astropy2018}
{Astropy Collaboration}, {Price-Whelan}, A.~M., {Sip{\H{o}}cz}, B.~M., {et~al.}
  2018, \href{https://doi.org/10.3847/1538-3881/aabc4f}{\aj},
  \href{https://ui.adsabs.harvard.edu/abs/2018AJ....156..123A}{156, 123}

\bibitem[{{Astropy Collaboration} {et~al.}(2013){Astropy Collaboration},
  {Robitaille}, {Tollerud}, {Greenfield}, {Droettboom}, {Bray}, {Aldcroft},
  {Davis}, {Ginsburg}, {Price-Whelan}, {Kerzendorf}, {Conley}, {Crighton},
  {Barbary}, {Muna}, {Ferguson}, {Grollier}, {Parikh}, {Nair}, {Unther},
  {Deil}, {Woillez}, {Conseil}, {Kramer}, {Turner}, {Singer}, {Fox}, {Weaver},
  {Zabalza}, {Edwards}, {Azalee Bostroem}, {Burke}, {Casey}, {Crawford},
  {Dencheva}, {Ely}, {Jenness}, {Labrie}, {Lim}, {Pierfederici}, {Pontzen},
  {Ptak}, {Refsdal}, {Servillat}, \& {Streicher}}]{astropy2013}
{Astropy Collaboration}, {Robitaille}, T.~P., {Tollerud}, E.~J., {et~al.} 2013,
  \href{https://doi.org/10.1051/0004-6361/201322068}{\aap},
  \href{https://ui.adsabs.harvard.edu/abs/2013A&A...558A..33A}{558, A33}

\bibitem[{{Atreya} {et~al.}(1999){Atreya}, {Wong}, {Owen}, {Mahaffy},
  {Niemann}, {de Pater}, {Drossart}, \& {Encrenaz}}]{atreya1999}
{Atreya}, S.~K., {Wong}, M.~H., {Owen}, T.~C., {et~al.} 1999,
  \href{https://doi.org/10.1016/S0032-0633(99)00047-1}{\planss},
  \href{https://ui.adsabs.harvard.edu/abs/1999P&SS...47.1243A}{47, 1243}

\bibitem[{{Bailey}(2007)}]{bailey2007}
{Bailey}, J. 2007, \href{https://doi.org/10.1089/ast.2006.0039}{Astrobiology},
  \href{https://ui.adsabs.harvard.edu/abs/2007AsBio...7..320B}{7, 320}

\bibitem[{{Bailey} {et~al.}(2021){Bailey}, {Bott}, {Cotton},
  {Kedziora-Chudczer}, {Zhao}, {Evensberget}, {Marshall}, {Wright}, \&
  {Lucas}}]{bailey2021}
{Bailey}, J., {Bott}, K., {Cotton}, D.~V., {et~al.} 2021,
  \href{https://doi.org/10.1093/mnras/stab172}{\mnras},
  \href{https://ui.adsabs.harvard.edu/abs/2021MNRAS.502.2331B}{502, 2331}

\bibitem[{{Bailey} {et~al.}(2020){Bailey}, {Cotton}, {Kedziora-Chudczer}, {De
  Horta}, \& {Maybour}}]{bailey2020}
{Bailey}, J., {Cotton}, D.~V., {Kedziora-Chudczer}, L., {De Horta}, A., \&
  {Maybour}, D. 2020, \href{https://doi.org/10.1017/pasa.2019.45}{\pasa},
  \href{https://ui.adsabs.harvard.edu/abs/2020PASA...37....4B}{37, e004}

\bibitem[{{Bailey} {et~al.}(2018){Bailey}, {Kedziora-Chudczer}, \&
  {Bott}}]{bailey2018}
{Bailey}, J., {Kedziora-Chudczer}, L., \& {Bott}, K. 2018,
  \href{https://doi.org/10.1093/mnras/sty1892}{\mnras},
  \href{https://ui.adsabs.harvard.edu/abs/2018MNRAS.480.1613B}{480, 1613}

\bibitem[{{Bailey} {et~al.}(2015){Bailey}, {Kedziora-Chudczer}, {Cotton},
  {Bott}, {Hough}, \& {Lucas}}]{bailey2015}
{Bailey}, J., {Kedziora-Chudczer}, L., {Cotton}, D.~V., {et~al.} 2015,
  \href{https://doi.org/10.1093/mnras/stv519}{\mnras},
  \href{https://ui.adsabs.harvard.edu/abs/2015MNRAS.449.3064B}{449, 3064}

\bibitem[{{Berdyugina} {et~al.}(2008){Berdyugina}, {Berdyugin}, {Fluri}, \&
  {Piirola}}]{berdyugina2008}
{Berdyugina}, S.~V., {Berdyugin}, A.~V., {Fluri}, D.~M., \& {Piirola}, V. 2008,
  \href{https://doi.org/10.1086/527320}{\apjl},
  \href{https://ui.adsabs.harvard.edu/abs/2008ApJ...673L..83B}{673, L83}

\bibitem[{{Berdyugina} {et~al.}(2011){Berdyugina}, {Berdyugin}, {Fluri}, \&
  {Piirola}}]{berdyugina2011}
{Berdyugina}, S.~V., {Berdyugin}, A.~V., {Fluri}, D.~M., \& {Piirola}, V. 2011,
  \href{https://doi.org/10.1088/2041-8205/728/1/L6}{\apjl},
  \href{https://ui.adsabs.harvard.edu/abs/2011ApJ...728L...6B}{728, L6}

\bibitem[{{Beuzit} {et~al.}(2019){Beuzit}, {Vigan}, {Mouillet}, {Dohlen},
  {Gratton}, {Boccaletti}, {Sauvage}, {Schmid}, {Langlois}, {Petit},
  {Baruffolo}, {Feldt}, {Milli}, {Wahhaj}, {Abe}, {Anselmi}, {Antichi},
  {Barette}, {Baudrand}, {Baudoz}, {Bazzon}, {Bernardi}, {Blanchard}, {Brast},
  {Bruno}, {Buey}, {Carbillet}, {Carle}, {Cascone}, {Chapron}, {Charton},
  {Chauvin}, {Claudi}, {Costille}, {De Caprio}, {de Boer}, {Delboulb{\'e}},
  {Desidera}, {Dominik}, {Downing}, {Dupuis}, {Fabron}, {Fantinel}, {Farisato},
  {Feautrier}, {Fedrigo}, {Fusco}, {Gigan}, {Ginski}, {Girard}, {Giro},
  {Gisler}, {Gluck}, {Gry}, {Henning}, {Hubin}, {Hugot}, {Incorvaia}, {Jaquet},
  {Kasper}, {Lagadec}, {Lagrange}, {Le Coroller}, {Le Mignant}, {Le Ruyet},
  {Lessio}, {Lizon}, {Llored}, {Lundin}, {Madec}, {Magnard}, {Marteaud},
  {Martinez}, {Maurel}, {M{\'e}nard}, {Mesa}, {M{\"o}ller-Nilsson}, {Moulin},
  {Moutou}, {Orign{\'e}}, {Parisot}, {Pavlov}, {Perret}, {Pragt}, {Puget},
  {Rabou}, {Ramos}, {Reess}, {Rigal}, {Rochat}, {Roelfsema}, {Rousset}, {Roux},
  {Saisse}, {Salasnich}, {Santambrogio}, {Scuderi}, {Segransan}, {Sevin},
  {Siebenmorgen}, {Soenke}, {Stadler}, {Suarez}, {Tiph{\`e}ne}, {Turatto},
  {Udry}, {Vakili}, {Waters}, {Weber}, {Wildi}, {Zins}, \&
  {Zurlo}}]{beuzit2019}
{Beuzit}, J.~L., {Vigan}, A., {Mouillet}, D., {et~al.} 2019,
  \href{https://doi.org/10.1051/0004-6361/201935251}{\aap},
  \href{https://ui.adsabs.harvard.edu/abs/2019A&A...631A.155B}{631, A155}

\bibitem[{{Bohren} \& {Huffman}(1983)}]{bohren1983}
{Bohren}, C.~F. \& {Huffman}, D.~R. 1983,
  \href{https://ui.adsabs.harvard.edu/abs/1983asls.book.....B}{{Absorption and
  scattering of light by small particles}} (John Wiley \& Sons)

\bibitem[{{Bott} {et~al.}(2018){Bott}, {Bailey}, {Cotton}, {Kedziora-Chudczer},
  {Marshall}, \& {Meadows}}]{bott2018}
{Bott}, K., {Bailey}, J., {Cotton}, D.~V., {et~al.} 2018,
  \href{https://doi.org/10.3847/1538-3881/aaed20}{\aj},
  \href{https://ui.adsabs.harvard.edu/abs/2018AJ....156..293B}{156, 293}

\bibitem[{{Bott} {et~al.}(2016){Bott}, {Bailey}, {Kedziora-Chudczer}, {Cotton},
  {Lucas}, {Marshall}, \& {Hough}}]{bott2016}
{Bott}, K., {Bailey}, J., {Kedziora-Chudczer}, L., {et~al.} 2016,
  \href{https://doi.org/10.1093/mnrasl/slw046}{\mnras},
  \href{https://ui.adsabs.harvard.edu/abs/2016MNRAS.459L.109B}{459, L109}

\bibitem[{{Bouchy} {et~al.}(2005){Bouchy}, {Udry}, {Mayor}, {Moutou}, {Pont},
  {Iribarne}, {da Silva}, {Ilovaisky}, {Queloz}, {Santos}, {S{\'e}gransan}, \&
  {Zucker}}]{bouchy2005}
{Bouchy}, F., {Udry}, S., {Mayor}, M., {et~al.} 2005,
  \href{https://doi.org/10.1051/0004-6361:200500201}{\aap},
  \href{https://ui.adsabs.harvard.edu/abs/2005A&A...444L..15B}{444, L15}

\bibitem[{{Boyajian} {et~al.}(2015){Boyajian}, {von Braun}, {Feiden}, {Huber},
  {Basu}, {Demarque}, {Fischer}, {Schaefer}, {Mann}, {White}, {Maestro},
  {Brewer}, {Lamell}, {Spada}, {L{\'o}pez-Morales}, {Ireland}, {Farrington},
  {van Belle}, {Kane}, {Jones}, {ten Brummelaar}, {Ciardi}, {McAlister},
  {Ridgway}, {Goldfinger}, {Turner}, \& {Sturmann}}]{boyajian2015}
{Boyajian}, T., {von Braun}, K., {Feiden}, G.~A., {et~al.} 2015,
  \href{https://doi.org/10.1093/mnras/stu2502}{\mnras},
  \href{https://ui.adsabs.harvard.edu/abs/2015MNRAS.447..846B}{447, 846}

\bibitem[{{Brauer} {et~al.}(2016){Brauer}, {Wolf}, \& {Reissl}}]{brauer2016}
{Brauer}, R., {Wolf}, S., \& {Reissl}, S. 2016,
  \href{https://doi.org/10.1051/0004-6361/201527546}{\aap},
  \href{https://ui.adsabs.harvard.edu/abs/2016A&A...588A.129B}{588, A129}

\bibitem[{{Brewster}(1815)}]{brewster1815}
{Brewster}, D. 1815, \href{https://doi.org/10.1098/rstl.1815.0010}{Phil. Trans.
  R. Soc.}, \href{https://ui.adsabs.harvard.edu/abs/1815RSPT..105..125B}{105,
  125}

\bibitem[{{Brunngr{\"a}ber} \& {Wolf}(2020)}]{brunngraeber2020}
{Brunngr{\"a}ber}, R. \& {Wolf}, S. 2020,
  \href{https://doi.org/10.1051/0004-6361/202037981}{\aap},
  \href{https://ui.adsabs.harvard.edu/abs/2020A&A...640A.122B}{640, A122}

\bibitem[{{Brunngr{\"a}ber} \& {Wolf}(2021)}]{brunngraeber2021}
{Brunngr{\"a}ber}, R. \& {Wolf}, S. 2021,
  \href{https://doi.org/10.1051/0004-6361/202040033}{\aap},
  \href{https://ui.adsabs.harvard.edu/abs/2021A&A...648A..87B}{648, A87}

\bibitem[{{Buenzli} \& {Schmid}(2009)}]{buenzli2009}
{Buenzli}, E. \& {Schmid}, H.~M. 2009,
  \href{https://doi.org/10.1051/0004-6361/200911760}{\aap},
  \href{https://ui.adsabs.harvard.edu/abs/2009A&A...504..259B}{504, 259}

\bibitem[{{Burrows} {et~al.}(1997){Burrows}, {Marley}, {Hubbard}, {Lunine},
  {Guillot}, {Saumon}, {Freedman}, {Sudarsky}, \& {Sharp}}]{burrows1997}
{Burrows}, A., {Marley}, M., {Hubbard}, W.~B., {et~al.} 1997,
  \href{https://doi.org/10.1086/305002}{\apj},
  \href{https://ui.adsabs.harvard.edu/abs/1997ApJ...491..856B}{491, 856}

\bibitem[{{Burrows} \& {Sharp}(1999)}]{burrows1999}
{Burrows}, A. \& {Sharp}, C.~M. 1999,
  \href{https://doi.org/10.1086/306811}{\apj},
  \href{https://ui.adsabs.harvard.edu/abs/1999ApJ...512..843B}{512, 843}

\bibitem[{{Carlson} {et~al.}(1988){Carlson}, {Rossow}, \&
  {Orton}}]{carlson1988}
{Carlson}, B.~E., {Rossow}, W.~B., \& {Orton}, G.~S. 1988,
  \href{https://doi.org/10.1175/1520-0469(1988)045<2066:CMOTGP>2.0.CO;2}{J.
  Atmos. Sci.},
  \href{https://ui.adsabs.harvard.edu/abs/1988JAtS...45.2066C}{45, 2066}

\bibitem[{{Charbonneau} {et~al.}(2002){Charbonneau}, {Brown}, {Noyes}, \&
  {Gilliland}}]{charbonneau2002}
{Charbonneau}, D., {Brown}, T.~M., {Noyes}, R.~W., \& {Gilliland}, R.~L. 2002,
  \href{https://doi.org/10.1086/338770}{\apj},
  \href{https://ui.adsabs.harvard.edu/abs/2002ApJ...568..377C}{568, 377}

\bibitem[{{Chernova} {et~al.}(1993){Chernova}, {Kiselev}, \&
  {Jockers}}]{chernova1993}
{Chernova}, G.~P., {Kiselev}, N.~N., \& {Jockers}, K. 1993,
  \href{https://doi.org/10.1006/icar.1993.1063}{\icarus},
  \href{https://ui.adsabs.harvard.edu/abs/1993Icar..103..144C}{103, 144}

\bibitem[{{Coffeen} \& {Gehrels}(1969)}]{coffeen1969}
{Coffeen}, D.~L. \& {Gehrels}, T. 1969,
  \href{https://doi.org/10.1086/110822}{\aj},
  \href{https://ui.adsabs.harvard.edu/abs/1969AJ.....74..433C}{74, 433}

\bibitem[{{Cotton} {et~al.}(2017){Cotton}, {Marshall}, {Bailey},
  {Kedziora-Chudczer}, {Bott}, {Marsden}, \& {Carter}}]{cotton2017}
{Cotton}, D.~V., {Marshall}, J.~P., {Bailey}, J., {et~al.} 2017,
  \href{https://doi.org/10.1093/mnras/stx068}{\mnras},
  \href{https://ui.adsabs.harvard.edu/abs/2017MNRAS.467..873C}{467, 873}

\bibitem[{{Cox}(2000)}]{cox2000}
{Cox}, A.~N. 2000,
  \href{https://ui.adsabs.harvard.edu/abs/2000asqu.book.....C}{{Allen's
  astrophysical quantities}} (Springer)

\bibitem[{{de Haan} {et~al.}(1987){de Haan}, {Bosma}, \&
  {Hovenier}}]{de-haan1987}
{de Haan}, J.~F., {Bosma}, P.~B., \& {Hovenier}, J.~W. 1987, \aap,
  \href{https://ui.adsabs.harvard.edu/abs/1987A&A...183..371D}{183, 371}

\bibitem[{{Demory} {et~al.}(2011){Demory}, {Seager}, {Madhusudhan}, {Kjeldsen},
  {Christensen-Dalsgaard}, {Gillon}, {Rowe}, {Welsh}, {Adams}, {Dupree},
  {McCarthy}, {Kulesa}, {Borucki}, \& {Koch}}]{demory2011}
{Demory}, B., {Seager}, S., {Madhusudhan}, N., {et~al.} 2011,
  \href{https://doi.org/10.1088/2041-8205/735/1/L12}{\apjl},
  \href{https://ui.adsabs.harvard.edu/abs/2011ApJ...735L..12D}{735, L12}

\bibitem[{{Dorschner} {et~al.}(1995){Dorschner}, {Begemann}, {Henning},
  {J{\"a}ger}, \& {Mutschke}}]{dorschner1995}
{Dorschner}, J., {Begemann}, B., {Henning}, {\relax Th}., {J{\"a}ger}, C., \&
  {Mutschke}, H. 1995, \aap,
  \href{https://ui.adsabs.harvard.edu/abs/1995A&A...300..503D}{300, 503}

\bibitem[{{Draine}(2003)}]{draine2003}
{Draine}, B.~T. 2003, \href{https://doi.org/10.1086/379118}{\apj},
  \href{https://ui.adsabs.harvard.edu/abs/2003ApJ...598.1017D}{598, 1017}

\bibitem[{{Emde} {et~al.}(2016){Emde}, {Buras-Schnell}, {Kylling}, {Mayer},
  {Gasteiger}, {Hamann}, {Kylling}, {Richter}, {Pause}, {Dowling}, \&
  {Bugliaro}}]{emde2016}
{Emde}, C., {Buras-Schnell}, R., {Kylling}, A., {et~al.} 2016,
  \href{https://doi.org/10.5194/gmd-9-1647-2016}{Geophys. Mod. Dev.},
  \href{https://ui.adsabs.harvard.edu/abs/2016GMD.....9.1647E}{9, 1647}

\bibitem[{{Fabian} {et~al.}(2001){Fabian}, {Henning}, {J{\"a}ger}, {Mutschke},
  {Dorschner}, \& {Wehrhan}}]{fabian2001}
{Fabian}, D., {Henning}, {\relax Th}., {J{\"a}ger}, C., {et~al.} 2001,
  \href{https://doi.org/10.1051/0004-6361:20011196}{\aap},
  \href{https://ui.adsabs.harvard.edu/abs/2001A&A...378..228F}{378, 228}

\bibitem[{{Garc{\'\i}a Mu{\~n}oz} \& {Mills}(2015)}]{garcia-munoz2015}
{Garc{\'\i}a Mu{\~n}oz}, A. \& {Mills}, F.~P. 2015,
  \href{https://doi.org/10.1051/0004-6361/201424042}{\aap},
  \href{https://ui.adsabs.harvard.edu/abs/2015A&A...573A..72G}{573, A72}

\bibitem[{{Garc{\'\i}a Mu{\~n}oz} {et~al.}(2014){Garc{\'\i}a Mu{\~n}oz},
  {P{\'e}rez-Hoyos}, \& {S{\'a}nchez-Lavega}}]{garcia-munoz2014}
{Garc{\'\i}a Mu{\~n}oz}, A., {P{\'e}rez-Hoyos}, S., \& {S{\'a}nchez-Lavega}, A.
  2014, \href{https://doi.org/10.1051/0004-6361/201423531}{\aap},
  \href{https://ui.adsabs.harvard.edu/abs/2014A&A...566L...1G}{566, L1}

\bibitem[{{Gibson} {et~al.}(2013){Gibson}, {Aigrain}, {Barstow}, {Evans},
  {Fletcher}, \& {Irwin}}]{gibson2013}
{Gibson}, N.~P., {Aigrain}, S., {Barstow}, J.~K., {et~al.} 2013,
  \href{https://doi.org/10.1093/mnras/stt1783}{\mnras},
  \href{https://ui.adsabs.harvard.edu/abs/2013MNRAS.436.2974G}{436, 2974}

\bibitem[{{Goloub} {et~al.}(1994){Goloub}, {Deuze}, {Herman}, \&
  {Fouquart}}]{goloub1994}
{Goloub}, P., {Deuze}, J.~L., {Herman}, M., \& {Fouquart}, Y. 1994,
  \href{https://doi.org/10.1109/36.285191}{IEEE Trans. Geosci. Remote Sens.},
  \href{https://ui.adsabs.harvard.edu/abs/1994ITGRS..32...78G}{32, 78}

\bibitem[{{Goloub} {et~al.}(2000){Goloub}, {Herman}, {Chepfer}, {Riedi},
  {Brogniez}, {Couvert}, \& {S{\'e}Ze}}]{goloub2000}
{Goloub}, P., {Herman}, M., {Chepfer}, H., {et~al.} 2000,
  \href{https://doi.org/10.1029/1999JD901183}{\jgr},
  \href{https://ui.adsabs.harvard.edu/abs/2000JGR...10514747G}{105, 14,747}

\bibitem[{{Guillot}(2010)}]{guillot2010}
{Guillot}, T. 2010, \href{https://doi.org/10.1051/0004-6361/200913396}{\aap},
  \href{https://ui.adsabs.harvard.edu/abs/2010A&A...520A..27G}{520, A27}

\bibitem[{{Hale} \& {Querry}(1973)}]{hale1973}
{Hale}, G.~M. \& {Querry}, M.~R. 1973,
  \href{https://doi.org/10.1364/AO.12.000555}{\ao},
  \href{https://ui.adsabs.harvard.edu/abs/1973ApOpt..12..555H}{12, 555}

\bibitem[{{Hansen}(1971)}]{hansen1971}
{Hansen}, J.~E. 1971,
  \href{https://doi.org/10.1175/1520-0469(1971)028<1400:MSOPLI>2.0.CO;2}{J.
  Atmos. Sci.},
  \href{https://ui.adsabs.harvard.edu/abs/1971JAtS...28.1400H}{28, 1400}

\bibitem[{{Hansen} \& {Hovenier}(1974)}]{hansen1974a}
{Hansen}, J.~E. \& {Hovenier}, J.~W. 1974,
  \href{https://doi.org/10.1175/1520-0469(1974)031<1137:IOTPOV>2.0.CO;2}{J.
  Atmos. Sci.},
  \href{https://ui.adsabs.harvard.edu/abs/1974JAtS...31.1137H}{31, 1137}

\bibitem[{{Hansen} \& {Travis}(1974)}]{hansen1974b}
{Hansen}, J.~E. \& {Travis}, L.~D. 1974,
  \href{https://doi.org/10.1007/BF00168069}{\ssr},
  \href{https://ui.adsabs.harvard.edu/abs/1974SSRv...16..527H}{16, 527}

\bibitem[{{Harris} {et~al.}(2020){Harris}, {Millman}, {van der Walt},
  {Gommers}, {Virtanen}, {Cournapeau}, {Wieser}, {Taylor}, {Berg}, {Smith},
  {Kern}, {Picus}, {Hoyer}, {van Kerkwijk}, {Brett}, {Haldane}, {del R{\'\i}o},
  {Wiebe}, {Peterson}, {G{\'e}rard-Marchant}, {Sheppard}, {Reddy}, {Weckesser},
  {Abbasi}, {Gohlke}, \& {Oliphant}}]{harris2020}
{Harris}, C.~R., {Millman}, K.~J., {van der Walt}, S.~J., {et~al.} 2020,
  \href{https://doi.org/10.1038/s41586-020-2649-2}{\nat},
  \href{https://ui.adsabs.harvard.edu/abs/2020Natur.585..357H}{585, 357}

\bibitem[{{Heese} {et~al.}(2020){Heese}, {Wolf}, \& {Brauer}}]{heese2020}
{Heese}, S., {Wolf}, S., \& {Brauer}, R. 2020,
  \href{https://doi.org/10.1051/0004-6361/201935377}{\aap},
  \href{https://ui.adsabs.harvard.edu/abs/2020A&A...634A.129H}{634, A129}

\bibitem[{{Hellier} {et~al.}(2009){Hellier}, {Anderson}, {Collier Cameron},
  {Gillon}, {Hebb}, {Maxted}, {Queloz}, {Smalley}, {Triaud}, {West}, {Wilson},
  {Bentley}, {Enoch}, {Horne}, {Irwin}, {Lister}, {Mayor}, {Parley}, {Pepe},
  {Pollacco}, {Segransan}, {Udry}, \& {Wheatley}}]{hellier2009}
{Hellier}, C., {Anderson}, D.~R., {Collier Cameron}, A., {et~al.} 2009,
  \href{https://doi.org/10.1038/nature08245}{\nat},
  \href{https://ui.adsabs.harvard.edu/abs/2009Natur.460.1098H}{460, 1098}

\bibitem[{{Helling} {et~al.}(2001){Helling}, {Oevermann}, {L{\"u}ttke},
  {Klein}, \& {Sedlmayr}}]{helling2001}
{Helling}, {\relax Ch}., {Oevermann}, M., {L{\"u}ttke}, M.~J.~H., {Klein}, R.,
  \& {Sedlmayr}, E. 2001,
  \href{https://doi.org/10.1051/0004-6361:20010937}{\aap},
  \href{https://ui.adsabs.harvard.edu/abs/2001A&A...376..194H}{376, 194}

\bibitem[{{Helling} {et~al.}(2008){Helling}, {Woitke}, \& {Thi}}]{helling2008}
{Helling}, {\relax Ch}., {Woitke}, P., \& {Thi}, W. 2008,
  \href{https://doi.org/10.1051/0004-6361:20078220}{\aap},
  \href{https://ui.adsabs.harvard.edu/abs/2008A&A...485..547H}{485, 547}

\bibitem[{{Henning} {et~al.}(1995){Henning}, {Begemann}, {Mutschke}, \&
  {Dorschner}}]{henning1995}
{Henning}, {\relax Th}., {Begemann}, B., {Mutschke}, H., \& {Dorschner}, J.
  1995, \aaps,
  \href{https://ui.adsabs.harvard.edu/abs/1995A&AS..112..143H}{112, 143}

\bibitem[{{Hess}(1998)}]{hess1998}
{Hess}, M. 1998, \href{https://doi.org/10.1016/S0022-4073(98)00007-7}{\jqsrt},
  \href{https://ui.adsabs.harvard.edu/abs/1998JQSRT..60..301H}{60, 301}

\bibitem[{{Huffman} \& {Wild}(1967)}]{huffman1967}
{Huffman}, D.~R. \& {Wild}, R.~L. 1967,
  \href{https://doi.org/10.1103/PhysRev.156.989}{Phys. Rev.},
  \href{https://ui.adsabs.harvard.edu/abs/1967PhRv..156..989H}{156, 989}

\bibitem[{{Hunter}(2007)}]{hunter2007}
{Hunter}, J.~D. 2007, \href{https://doi.org/10.1109/MCSE.2007.55}{Comput. Sci.
  Eng.}, \href{https://ui.adsabs.harvard.edu/abs/2007CSE.....9...90H}{9, 90}

\bibitem[{{Hunziker} {et~al.}(2020){Hunziker}, {Schmid}, {Mouillet}, {Milli},
  {Zurlo}, {Delorme}, {Abe}, {Avenhaus}, {Baruffolo}, {Bazzon}, {Boccaletti},
  {Baudoz}, {Beuzit}, {Carbillet}, {Chauvin}, {Claudi}, {Costille}, {Daban},
  {Desidera}, {Dohlen}, {Dominik}, {Downing}, {Engler}, {Feldt}, {Fusco},
  {Ginski}, {Gisler}, {Girard}, {Gratton}, {Henning}, {Hubin}, {Kasper},
  {Keller}, {Langlois}, {Lagadec}, {Martinez}, {Maire}, {Menard}, {Meyer},
  {Pavlov}, {Pragt}, {Puget}, {Quanz}, {Rickman}, {Roelfsema}, {Salasnich},
  {Sauvage}, {Siebenmorgen}, {Sissa}, {Snik}, {Suarez}, {Szul{\'a}gyi},
  {Thalmann}, {Turatto}, {Udry}, {van Holstein}, {Vigan}, \&
  {Wildi}}]{hunziker2020}
{Hunziker}, S., {Schmid}, H.~M., {Mouillet}, D., {et~al.} 2020,
  \href{https://doi.org/10.1051/0004-6361/201936641}{\aap},
  \href{https://ui.adsabs.harvard.edu/abs/2020A&A...634A..69H}{634, A69}

\bibitem[{{J{\"a}ger} {et~al.}(2003){J{\"a}ger}, {Dorschner}, {Mutschke},
  {Posch}, \& {Henning}}]{jaeger2003}
{J{\"a}ger}, C., {Dorschner}, J., {Mutschke}, H., {Posch}, T., \& {Henning},
  {\relax Th}. 2003, \href{https://doi.org/10.1051/0004-6361:20030916}{\aap},
  \href{https://ui.adsabs.harvard.edu/abs/2003A&A...408..193J}{408, 193}

\bibitem[{{Karalidi} \& {Stam}(2012)}]{karalidi2012a}
{Karalidi}, T. \& {Stam}, D.~M. 2012,
  \href{https://doi.org/10.1051/0004-6361/201219297}{\aap},
  \href{https://ui.adsabs.harvard.edu/abs/2012A&A...546A..56K}{546, A56}

\bibitem[{{Karalidi} {et~al.}(2013){Karalidi}, {Stam}, \&
  {Guirado}}]{karalidi2013}
{Karalidi}, T., {Stam}, D.~M., \& {Guirado}, D. 2013,
  \href{https://doi.org/10.1051/0004-6361/201321492}{\aap},
  \href{https://ui.adsabs.harvard.edu/abs/2013A&A...555A.127K}{555, A127}

\bibitem[{{Karalidi} {et~al.}(2011){Karalidi}, {Stam}, \&
  {Hovenier}}]{karalidi2011}
{Karalidi}, T., {Stam}, D.~M., \& {Hovenier}, J.~W. 2011,
  \href{https://doi.org/10.1051/0004-6361/201116449}{\aap},
  \href{https://ui.adsabs.harvard.edu/abs/2011A&A...530A..69K}{530, A69}

\bibitem[{{Karalidi} {et~al.}(2012){Karalidi}, {Stam}, \&
  {Hovenier}}]{karalidi2012b}
{Karalidi}, T., {Stam}, D.~M., \& {Hovenier}, J.~W. 2012,
  \href{https://doi.org/10.1051/0004-6361/201220245}{\aap},
  \href{https://ui.adsabs.harvard.edu/abs/2012A&A...548A..90K}{548, A90}

\bibitem[{{Kemp} {et~al.}(1987){Kemp}, {Henson}, {Steiner}, \&
  {Powell}}]{kemp1987}
{Kemp}, J.~C., {Henson}, G.~D., {Steiner}, C.~T., \& {Powell}, E.~R. 1987,
  \href{https://doi.org/10.1038/326270a0}{\nat},
  \href{https://ui.adsabs.harvard.edu/abs/1987Natur.326..270K}{326, 270}

\bibitem[{{Khachai} {et~al.}(2009){Khachai}, {Khenata}, {Bouhemadou}, {Haddou},
  {Reshak}, {Amrani}, {Rached}, \& {Soudini}}]{khachai2009}
{Khachai}, H., {Khenata}, R., {Bouhemadou}, A., {et~al.} 2009,
  \href{https://doi.org/10.1088/0953-8984/21/9/095404}{J. Phys. Condens.
  Matter}, \href{https://ui.adsabs.harvard.edu/abs/2009JPCM...21i5404K}{21,
  095404}

\bibitem[{{Kitzmann} \& {Heng}(2018)}]{kitzmann2018}
{Kitzmann}, D. \& {Heng}, K. 2018,
  \href{https://doi.org/10.1093/mnras/stx3141}{\mnras},
  \href{https://ui.adsabs.harvard.edu/abs/2018MNRAS.475...94K}{475, 94}

\bibitem[{{Kitzmann} {et~al.}(2010){Kitzmann}, {Patzer}, {von Paris}, {Godolt},
  {Stracke}, {Gebauer}, {Grenfell}, \& {Rauer}}]{kitzmann2010}
{Kitzmann}, D., {Patzer}, A.~B.~C., {von Paris}, P., {et~al.} 2010,
  \href{https://doi.org/10.1051/0004-6361/200913491}{\aap},
  \href{https://ui.adsabs.harvard.edu/abs/2010A&A...511A..66K}{511, A66}

\bibitem[{{Knibbe} {et~al.}(1997){Knibbe}, {de Haan}, {Hovenier}, \&
  {Travis}}]{knibbe1997}
{Knibbe}, W.~J.~J., {de Haan}, J.~F., {Hovenier}, J.~W., \& {Travis}, L.~D.
  1997, \href{https://doi.org/10.1029/97JE00312}{\jgr},
  \href{https://ui.adsabs.harvard.edu/abs/1997JGR...10210945K}{102, 10945}

\bibitem[{{Koike} {et~al.}(1995){Koike}, {Kaito}, {Yamamoto}, {Shibai},
  {Kimura}, \& {Suto}}]{koike1995}
{Koike}, C., {Kaito}, C., {Yamamoto}, T., {et~al.} 1995,
  \href{https://doi.org/10.1006/icar.1995.1055}{\icarus},
  \href{https://ui.adsabs.harvard.edu/abs/1995Icar..114..203K}{114, 203}

\bibitem[{{Kopparla} {et~al.}(2016){Kopparla}, {Natraj}, {Zhang}, {Swain},
  {Wiktorowicz}, \& {Yung}}]{kopparla2016}
{Kopparla}, P., {Natraj}, V., {Zhang}, {\relax Xi}., {et~al.} 2016,
  \href{https://doi.org/10.3847/0004-637X/817/1/32}{\apj},
  \href{https://ui.adsabs.harvard.edu/abs/2016ApJ...817...32K}{817, 32}

\bibitem[{{Lawson} {et~al.}(1998){Lawson}, {Heymsfield}, {Aulenbach}, \&
  {Jensen}}]{lawson1998}
{Lawson}, R.~P., {Heymsfield}, A.~J., {Aulenbach}, S.~M., \& {Jensen}, T.~L.
  1998, \href{https://doi.org/10.1029/98GL00241}{\grl},
  \href{https://ui.adsabs.harvard.edu/abs/1998GeoRL..25.1331L}{25, 1331}

\bibitem[{{Lee} {et~al.}(2016){Lee}, {Dobbs-Dixon}, {Helling}, {Bognar}, \&
  {Woitke}}]{lee2016}
{Lee}, E., {Dobbs-Dixon}, I., {Helling}, {\relax Ch}., {Bognar}, K., \&
  {Woitke}, P. 2016, \href{https://doi.org/10.1051/0004-6361/201628606}{\aap},
  \href{https://ui.adsabs.harvard.edu/abs/2016A&A...594A..48L}{594, A48}

\bibitem[{{Lee} {et~al.}(2015){Lee}, {Helling}, {Dobbs-Dixon}, \&
  {Juncher}}]{lee2015}
{Lee}, E., {Helling}, {\relax Ch}., {Dobbs-Dixon}, I., \& {Juncher}, D. 2015,
  \href{https://doi.org/10.1051/0004-6361/201525982}{\aap},
  \href{https://ui.adsabs.harvard.edu/abs/2015A&A...580A..12L}{580, A12}

\bibitem[{{Lietzow} {et~al.}(2021){Lietzow}, {Wolf}, \&
  {Brunngr{\"a}ber}}]{lietzow2021}
{Lietzow}, M., {Wolf}, S., \& {Brunngr{\"a}ber}, R. 2021,
  \href{https://doi.org/10.1051/0004-6361/202038932}{\aap},
  \href{https://ui.adsabs.harvard.edu/abs/2021A&A...645A.146L}{645, A146}

\bibitem[{{Lindal}(1992)}]{lindal1992}
{Lindal}, G.~F. 1992, \href{https://doi.org/10.1086/116119}{\aj},
  \href{https://ui.adsabs.harvard.edu/abs/1992AJ....103..967L}{103, 967}

\bibitem[{{Lindal} {et~al.}(1987){Lindal}, {Lyons}, {Sweetnam}, {Eshleman},
  {Hinson}, \& {Tyler}}]{lindal1987}
{Lindal}, G.~F., {Lyons}, J.~R., {Sweetnam}, D.~N., {et~al.} 1987,
  \href{https://doi.org/10.1029/JA092iA13p14987}{\jgr},
  \href{https://ui.adsabs.harvard.edu/abs/1987JGR....9214987L}{92, 14987}

\bibitem[{{Lindal} {et~al.}(1990){Lindal}, {Lyons}, {Sweetnam}, {Eshleman},
  {Hinson}, \& {Tyler}}]{lindal1990}
{Lindal}, G.~F., {Lyons}, J.~R., {Sweetnam}, D.~N., {et~al.} 1990,
  \href{https://doi.org/10.1029/GL017i010p01733}{\grl},
  \href{https://ui.adsabs.harvard.edu/abs/1990GeoRL..17.1733L}{17, 1733}

\bibitem[{{Liou} \& {Hansen}(1971)}]{liou1971}
{Liou}, K. \& {Hansen}, J.~E. 1971,
  \href{https://doi.org/10.1175/1520-0469(1971)028<0995:IAPFSS>2.0.CO;2}{J.
  Atmos. Sci.},
  \href{https://ui.adsabs.harvard.edu/abs/1971JAtS...28..995L}{28, 995}

\bibitem[{{Lodders}(1999)}]{lodders1999}
{Lodders}, K. 1999, \href{https://doi.org/10.1086/307387}{\apj},
  \href{https://ui.adsabs.harvard.edu/abs/1999ApJ...519..793L}{519, 793}

\bibitem[{{Lodders}(2003)}]{lodders2003}
{Lodders}, K. 2003, \href{https://doi.org/10.1086/375492}{\apj},
  \href{https://ui.adsabs.harvard.edu/abs/2003ApJ...591.1220L}{591, 1220}

\bibitem[{{Lodders} \& {Fegley}(2002)}]{lodders2002}
{Lodders}, K. \& {Fegley}, B. 2002,
  \href{https://doi.org/10.1006/icar.2001.6740}{\icarus},
  \href{https://ui.adsabs.harvard.edu/abs/2002Icar..155..393L}{155, 393}

\bibitem[{{Lyot}(1929)}]{lyot1929}
{Lyot}, B. 1929, Ann. Obs. Meudon, 8, 1

\bibitem[{{Marcoux}(1969)}]{marcoux1969}
{Marcoux}, J.~E. 1969, \href{https://doi.org/10.1364/JOSA.59.000998}{J. Opt.
  Soc. Am.}, 59, 998

\bibitem[{{Martonchik} \& {Orton}(1994)}]{martonchik1994}
{Martonchik}, J.~V. \& {Orton}, G.~S. 1994,
  \href{https://doi.org/10.1364/AO.33.008306}{\ao},
  \href{https://ui.adsabs.harvard.edu/abs/1994ApOpt..33.8306M}{33, 8306}

\bibitem[{{Martonchik} {et~al.}(1984){Martonchik}, {Orton}, \&
  {Appleby}}]{martonchik1984}
{Martonchik}, J.~V., {Orton}, G.~S., \& {Appleby}, J.~F. 1984,
  \href{https://doi.org/10.1364/AO.23.000541}{\ao},
  \href{https://ui.adsabs.harvard.edu/abs/1984ApOpt..23..541M}{23, 541}

\bibitem[{{Mie}(1908)}]{mie1908}
{Mie}, G. 1908, \href{https://doi.org/10.1002/andp.19083300302}{Ann. Phys.},
  \href{https://ui.adsabs.harvard.edu/abs/1908AnP...330..377M}{330, 377}

\bibitem[{{Mishchenko} \& {Travis}(1994)}]{mishchenko1994}
{Mishchenko}, M. \& {Travis}, L.~D. 1994,
  \href{https://doi.org/10.1364/AO.33.007206}{\ao},
  \href{https://ui.adsabs.harvard.edu/abs/1994ApOpt..33.7206M}{33, 7206}

\bibitem[{{Morley} {et~al.}(2013){Morley}, {Fortney}, {Kempton}, {Marley},
  {Visscher}, \& {Zahnle}}]{morley2013}
{Morley}, C.~V., {Fortney}, J.~J., {Kempton}, E.~M.~R., {et~al.} 2013,
  \href{https://doi.org/10.1088/0004-637X/775/1/33}{\apj},
  \href{https://ui.adsabs.harvard.edu/abs/2013ApJ...775...33M}{775, 33}

\bibitem[{{Morley} {et~al.}(2012){Morley}, {Fortney}, {Marley}, {Visscher},
  {Saumon}, \& {Leggett}}]{morley2012}
{Morley}, C.~V., {Fortney}, J.~J., {Marley}, M.~S., {et~al.} 2012,
  \href{https://doi.org/10.1088/0004-637X/756/2/172}{\apj},
  \href{https://ui.adsabs.harvard.edu/abs/2012ApJ...756..172M}{756, 172}

\bibitem[{{Morley} {et~al.}(2014){Morley}, {Marley}, {Fortney}, {Lupu},
  {Saumon}, {Greene}, \& {Lodders}}]{morley2014}
{Morley}, C.~V., {Marley}, M.~S., {Fortney}, J.~J., {et~al.} 2014,
  \href{https://doi.org/10.1088/0004-637X/787/1/78}{\apj},
  \href{https://ui.adsabs.harvard.edu/abs/2014ApJ...787...78M}{787, 78}

\bibitem[{{Muslimov} {et~al.}(2018){Muslimov}, {Bouret}, {Neiner}, {L{\'o}pez
  Ariste}, {Ferrari}, {Viv{\`e}s}, {Hugot}, {Grange}, {Lombardo}, {Lopes},
  {Costeraste}, \& {Brachet}}]{muslimov2018}
{Muslimov}, E., {Bouret}, J., {Neiner}, C., {et~al.} 2018,
  \href{https://doi.org/10.1117/12.2310133}{Proc. SPIE},
  \href{https://ui.adsabs.harvard.edu/abs/2018SPIE10699E..06M}{10699, 1069906}

\bibitem[{{Nussenzveig}(1969)}]{nussenzveig1969}
{Nussenzveig}, H.~M. 1969, \href{https://doi.org/10.1063/1.1664747}{J. Math.
  Phys.}, \href{https://ui.adsabs.harvard.edu/abs/1969JMP....10..125N}{10, 125}

\bibitem[{{Palik}(1985)}]{palik1985}
{Palik}, E.~D. 1985,
  \href{https://ui.adsabs.harvard.edu/abs/1985hocs.book.....P}{{Handbook of
  optical constants of solids}} (Academic Press)

\bibitem[{{Palik}(1991)}]{palik1991}
{Palik}, E.~D. 1991,
  \href{https://ui.adsabs.harvard.edu/abs/1991hocs.book.....P}{{Handbook of
  optical constants of solids II}} (Academic Press)

\bibitem[{{Pellegrini} {et~al.}(2020){Pellegrini}, {Reissl}, {Rahner},
  {Klessen}, {Glover}, {Pakmor}, {Herrera-Camus}, \& {Grand}}]{pellegrini2020}
{Pellegrini}, E.~W., {Reissl}, S., {Rahner}, D., {et~al.} 2020,
  \href{https://doi.org/10.1093/mnras/staa2555}{\mnras},
  \href{https://ui.adsabs.harvard.edu/abs/2020MNRAS.498.3193P}{498, 3193}

\bibitem[{{P{\'e}rez-Hoyos} {et~al.}(2005){P{\'e}rez-Hoyos},
  {S{\'a}nchez-Lavega}, {French}, \& {Rojas}}]{perez-hoyos2005}
{P{\'e}rez-Hoyos}, S., {S{\'a}nchez-Lavega}, A., {French}, R.~G., \& {Rojas},
  J.~F. 2005, \href{https://doi.org/10.1016/j.icarus.2005.01.014}{\icarus},
  \href{https://ui.adsabs.harvard.edu/abs/2005Icar..176..155P}{176, 155}

\bibitem[{{Perry} {et~al.}(1978){Perry}, {Huffman}, \& {Hunt}}]{perry1978}
{Perry}, R.~J., {Huffman}, D.~R., \& {Hunt}, A.~J. 1978,
  \href{https://doi.org/10.1364/AO.17.002700}{\ao},
  \href{https://ui.adsabs.harvard.edu/abs/1978ApOpt..17.2700P}{17, 2700}

\bibitem[{{Pollack} {et~al.}(1994){Pollack}, {Hollenbach}, {Beckwith},
  {Simonelli}, {Roush}, \& {Fong}}]{pollack1994}
{Pollack}, J.~B., {Hollenbach}, D., {Beckwith}, S., {et~al.} 1994,
  \href{https://doi.org/10.1086/173677}{\apj},
  \href{https://ui.adsabs.harvard.edu/abs/1994ApJ...421..615P}{421, 615}

\bibitem[{{Pont} {et~al.}(2008){Pont}, {Knutson}, {Gilliland}, {Moutou}, \&
  {Charbonneau}}]{pont2008}
{Pont}, F., {Knutson}, H., {Gilliland}, R.~L., {Moutou}, C., \& {Charbonneau},
  D. 2008, \href{https://doi.org/10.1111/j.1365-2966.2008.12852.x}{\mnras},
  \href{https://ui.adsabs.harvard.edu/abs/2008MNRAS.385..109P}{385, 109}

\bibitem[{{Reissl} {et~al.}(2017){Reissl}, {Seifried}, {Wolf}, {Banerjee}, \&
  {Klessen}}]{reissl2017}
{Reissl}, S., {Seifried}, D., {Wolf}, S., {Banerjee}, R., \& {Klessen}, R.~S.
  2017, \href{https://doi.org/10.1051/0004-6361/201730408}{\aap},
  \href{https://ui.adsabs.harvard.edu/abs/2017A&A...603A..71R}{603, A71}

\bibitem[{{Reissl} {et~al.}(2016){Reissl}, {Wolf}, \& {Brauer}}]{reissl2016}
{Reissl}, S., {Wolf}, S., \& {Brauer}, R. 2016,
  \href{https://doi.org/10.1051/0004-6361/201424930}{\aap},
  \href{https://ui.adsabs.harvard.edu/abs/2016A&A...593A..87R}{593, A87}

\bibitem[{{Rossi} {et~al.}(2018){Rossi}, {Berzosa-Molina}, \&
  {Stam}}]{rossi2018b}
{Rossi}, L., {Berzosa-Molina}, J., \& {Stam}, D.~M. 2018,
  \href{https://doi.org/10.1051/0004-6361/201832859}{\aap},
  \href{https://ui.adsabs.harvard.edu/abs/2018A&A...616A.147R}{616, A147}

\bibitem[{{Rossi} \& {Stam}(2017)}]{rossi2017}
{Rossi}, L. \& {Stam}, D.~M. 2017,
  \href{https://doi.org/10.1051/0004-6361/201730586}{\aap},
  \href{https://ui.adsabs.harvard.edu/abs/2017A&A...607A..57R}{607, A57}

\bibitem[{{Rossi} \& {Stam}(2018)}]{rossi2018a}
{Rossi}, L. \& {Stam}, D.~M. 2018,
  \href{https://doi.org/10.1051/0004-6361/201832619}{\aap},
  \href{https://ui.adsabs.harvard.edu/abs/2018A&A...616A.117R}{616, A117}

\bibitem[{{S{\'a}nchez-Lavega} {et~al.}(2004){S{\'a}nchez-Lavega},
  {P{\'e}rez-Hoyos}, \& {Hueso}}]{sanchez-lavega2004}
{S{\'a}nchez-Lavega}, A., {P{\'e}rez-Hoyos}, S., \& {Hueso}, R. 2004,
  \href{https://doi.org/10.1119/1.1645279}{Am. J. Phys.},
  \href{https://ui.adsabs.harvard.edu/abs/2004AmJPh..72..767S}{72, 767}

\bibitem[{{Sato} \& {Hansen}(1979)}]{sato1979}
{Sato}, M. \& {Hansen}, J.~E. 1979,
  \href{https://doi.org/10.1175/1520-0469(1979)036<1133:JACACS>2.0.CO;2}{J.
  Atmos. Sci.},
  \href{https://ui.adsabs.harvard.edu/abs/1979JAtS...36.1133S}{36, 1133}

\bibitem[{{Seager} {et~al.}(2000){Seager}, {Whitney}, \&
  {Sasselov}}]{seager2000}
{Seager}, S., {Whitney}, B.~A., \& {Sasselov}, D.~D. 2000,
  \href{https://doi.org/10.1086/309292}{\apj},
  \href{https://ui.adsabs.harvard.edu/abs/2000ApJ...540..504S}{540, 504}

\bibitem[{{Seifried} {et~al.}(2020){Seifried}, {Walch}, {Weis}, {Reissl},
  {Soler}, {Klessen}, \& {Joshi}}]{seifried2020}
{Seifried}, D., {Walch}, S., {Weis}, M., {et~al.} 2020,
  \href{https://doi.org/10.1093/mnras/staa2231}{\mnras},
  \href{https://ui.adsabs.harvard.edu/abs/2020MNRAS.497.4196S}{497, 4196}

\bibitem[{{Siefke} {et~al.}(2016){Siefke}, {Kroker}, {Pfeiffer}, {Puffky},
  {Dietrich}, {Franta}, {Ohl{\'\i}dal}, {Szeghalmi}, {Kley}, \&
  {T{\"u}nnermann}}]{siefke2016}
{Siefke}, T., {Kroker}, S., {Pfeiffer}, K., {et~al.} 2016,
  \href{https://doi.org/10.1002/adom.201600250}{Adv. Opt. Mater.},
  \href{https://ui.adsabs.harvard.edu/abs/2016arXiv160704866S}{4, 1780}

\bibitem[{{Sing} {et~al.}(2009){Sing}, {D{\'e}sert}, {Lecavelier Des Etangs},
  {Ballester}, {Vidal-Madjar}, {Parmentier}, {Hebrard}, \& {Henry}}]{sing2009}
{Sing}, D.~K., {D{\'e}sert}, J.~M., {Lecavelier Des Etangs}, A., {et~al.} 2009,
  \href{https://doi.org/10.1051/0004-6361/200912776}{\aap},
  \href{https://ui.adsabs.harvard.edu/abs/2009A&A...505..891S}{505, 891}

\bibitem[{{Sneep} \& {Ubachs}(2005)}]{sneep2005}
{Sneep}, M. \& {Ubachs}, W. 2005,
  \href{https://doi.org/10.1016/j.jqsrt.2004.07.025}{\jqsrt},
  \href{https://ui.adsabs.harvard.edu/abs/2005JQSRT..92..293S}{92, 293}

\bibitem[{{Snellen} {et~al.}(2010){Snellen}, {de Kok}, {de Mooij}, \&
  {Albrecht}}]{snellen2010}
{Snellen}, I.~A.~G., {de Kok}, R.~J., {de Mooij}, E.~J.~W., \& {Albrecht}, S.
  2010, \href{https://doi.org/10.1038/nature09111}{\nat},
  \href{https://ui.adsabs.harvard.edu/abs/2010Natur.465.1049S}{465, 1049}

\bibitem[{{Stam}(2008)}]{stam2008}
{Stam}, D.~M. 2008, \href{https://doi.org/10.1051/0004-6361:20078358}{\aap},
  \href{https://ui.adsabs.harvard.edu/abs/2008A&A...482..989S}{482, 989}

\bibitem[{{Stam} {et~al.}(2006){Stam}, {de Rooij}, {Cornet}, \&
  {Hovenier}}]{stam2006}
{Stam}, D.~M., {de Rooij}, W.~A., {Cornet}, G., \& {Hovenier}, J.~W. 2006,
  \href{https://doi.org/10.1051/0004-6361:20054364}{\aap},
  \href{https://ui.adsabs.harvard.edu/abs/2006A&A...452..669S}{452, 669}

\bibitem[{{Stam} {et~al.}(2004){Stam}, {Hovenier}, \& {Waters}}]{stam2004}
{Stam}, D.~M., {Hovenier}, J.~W., \& {Waters}, L.~B.~F.~M. 2004,
  \href{https://doi.org/10.1051/0004-6361:20041578}{\aap},
  \href{https://ui.adsabs.harvard.edu/abs/2004A&A...428..663S}{428, 663}

\bibitem[{{Stolker} {et~al.}(2017){Stolker}, {Min}, {Stam}, {Molli{\`e}re},
  {Dominik}, \& {Waters}}]{stolker2017}
{Stolker}, T., {Min}, M., {Stam}, D.~M., {et~al.} 2017,
  \href{https://doi.org/10.1051/0004-6361/201730780}{\aap},
  \href{https://ui.adsabs.harvard.edu/abs/2017A&A...607A..42S}{607, A42}

\bibitem[{{Ueda} {et~al.}(1998){Ueda}, {Yanagi}, {Noshiro}, {Hosono}, \&
  {Kawazoe}}]{ueda1998}
{Ueda}, K., {Yanagi}, H., {Noshiro}, R., {Hosono}, H., \& {Kawazoe}, H. 1998,
  \href{https://doi.org/10.1088/0953-8984/10/16/018}{J. Phys. Condens. Matter},
  \href{https://ui.adsabs.harvard.edu/abs/1998JPCM...10.3669U}{10, 3669}

\bibitem[{{van de Hulst}(1957)}]{van-de-hulst1957}
{van de Hulst}, H.~C. 1957,
  \href{https://ui.adsabs.harvard.edu/abs/1957lssp.book.....V}{{Light
  Scattering by Small Particles}} (John Wiley \& Sons)

\bibitem[{{Visscher} {et~al.}(2006){Visscher}, {Lodders}, \&
  {Fegley}}]{visscher2006}
{Visscher}, C., {Lodders}, K., \& {Fegley}, B. 2006,
  \href{https://doi.org/10.1086/506245}{\apj},
  \href{https://ui.adsabs.harvard.edu/abs/2006ApJ...648.1181V}{648, 1181}

\bibitem[{{Visscher} {et~al.}(2010){Visscher}, {Lodders}, \&
  {Fegley}}]{visscher2010}
{Visscher}, C., {Lodders}, K., \& {Fegley}, B. 2010,
  \href{https://doi.org/10.1088/0004-637X/716/2/1060}{\apj},
  \href{https://ui.adsabs.harvard.edu/abs/2010ApJ...716.1060V}{716, 1060}

\bibitem[{{Wakeford} \& {Sing}(2015)}]{wakeford2015}
{Wakeford}, H.~R. \& {Sing}, D.~K. 2015,
  \href{https://doi.org/10.1051/0004-6361/201424207}{\aap},
  \href{https://ui.adsabs.harvard.edu/abs/2015A&A...573A.122W}{573, A122}

\bibitem[{{Warren} \& {Brandt}(2008)}]{warren2008}
{Warren}, S.~G. \& {Brandt}, R.~E. 2008,
  \href{https://doi.org/10.1029/2007JD009744}{J. Geophys. Res.},
  \href{https://ui.adsabs.harvard.edu/abs/2008JGRD..11314220W}{113, D14220}

\bibitem[{{West} {et~al.}(1986){West}, {Strobel}, \& {Tomasko}}]{west1986}
{West}, R.~A., {Strobel}, D.~F., \& {Tomasko}, M.~G. 1986,
  \href{https://doi.org/10.1016/0019-1035(86)90135-1}{\icarus},
  \href{https://ui.adsabs.harvard.edu/abs/1986Icar...65..161W}{65, 161}

\bibitem[{{Wiktorowicz} \& {Nofi}(2015)}]{wiktorowicz2015}
{Wiktorowicz}, S.~J. \& {Nofi}, L.~A. 2015,
  \href{https://doi.org/10.1088/2041-8205/800/1/L1}{\apjl},
  \href{https://ui.adsabs.harvard.edu/abs/2015ApJ...800L...1W}{800, L1}

\bibitem[{{Wolf} \& {Voshchinnikov}(2004)}]{wolf2004}
{Wolf}, S. \& {Voshchinnikov}, N.~V. 2004,
  \href{https://doi.org/10.1016/j.cpc.2004.06.070}{Comput. Phys. Commun.},
  \href{https://ui.adsabs.harvard.edu/abs/2004CoPhC.162..113W}{162, 113}

\bibitem[{{Young}(1973)}]{young1973}
{Young}, A.~T. 1973,
  \href{https://doi.org/10.1016/0019-1035(73)90059-6}{\icarus},
  \href{https://ui.adsabs.harvard.edu/abs/1973Icar...18..564Y}{18, 564}

\bibitem[{{Zeidler} {et~al.}(2011){Zeidler}, {Posch}, {Mutschke}, {Richter}, \&
  {Wehrhan}}]{zeidler2011}
{Zeidler}, S., {Posch}, T., {Mutschke}, H., {Richter}, H., \& {Wehrhan}, O.
  2011, \href{https://doi.org/10.1051/0004-6361/201015219}{\aap},
  \href{https://ui.adsabs.harvard.edu/abs/2011A&A...526A..68Z}{526, A68}

\bibitem[{{Zielinski} {et~al.}(2021){Zielinski}, {Wolf}, \&
  {Brunngr{\"a}ber}}]{zielinski2021}
{Zielinski}, N., {Wolf}, S., \& {Brunngr{\"a}ber}, R. 2021,
  \href{https://doi.org/10.1051/0004-6361/202039126}{\aap},
  \href{https://ui.adsabs.harvard.edu/abs/2021A&A...645A.125Z}{645, A125}

\end{thebibliography}

\end{document}